\documentclass{aa}

\usepackage{graphicx}
\usepackage[T1]{fontenc}
\usepackage[usenames,dvipsnames]{xcolor}
\usepackage{natbib}
\usepackage[normalem]{ulem}
\usepackage{csvsimple}
\usepackage{xtab,afterpage}
\usepackage{txfonts}
\usepackage[backref=page, hyperfootnotes=true, hidelinks, colorlinks, citecolor=blue, linkcolor=blue, linktocpage, bookmarks=true]{hyperref}
\usepackage{footnote}
\usepackage{float}
\usepackage{tabularx}
\usepackage{threeparttablex}
\usepackage{enumitem}
\usepackage[bottom, flushmargin]{footmisc}

\newcolumntype{Y}{>{\centering\arraybackslash}X}

\makeatletter
\newbox\LT@firstfoot
\def\endfirstfoot{\LT@end@hd@ft\LT@firstfoot}
\newdimen\LT@footdiff
\def\LT@start{%
  \let\LT@start\endgraf
  \endgraf\penalty\z@
  \vskip\LTpre\endgraf
  \LT@footdiff-\ht\LT@foot
  \advance\LT@footdiff\ht\LT@firstfoot
  \dimen@\pagetotal
  \advance\dimen@ \ht\ifvoid\LT@firsthead\LT@head\else\LT@firsthead\fi
  \advance\dimen@ \dp\ifvoid\LT@firsthead\LT@head\else\LT@firsthead\fi
  \advance\dimen@ \ht\ifvoid\LT@firstfoot\LT@foot\else\LT@firstfoot\fi
  \dimen@ii\vfuzz
  \vfuzz\maxdimen
  \setbox\tw@\copy\z@
  \setbox\tw@\vsplit\tw@ to \ht\@arstrutbox
  \setbox\tw@\vbox{\unvbox\tw@}%
  \vfuzz\dimen@ii
  \advance\dimen@ \ht
      \ifdim\ht\@arstrutbox>\ht\tw@\@arstrutbox\else\tw@\fi
  \advance\dimen@\dp
      \ifdim\dp\@arstrutbox>\dp\tw@\@arstrutbox\else\tw@\fi
  \advance\dimen@ -\pagegoal
  \ifdim \dimen@>\z@\vfil\break\fi
  \global\@colroom\@colht
  \ifvoid\LT@firstfoot
    \ifvoid\LT@foot
    \else
      \advance\vsize-\ht\LT@foot
      \global\advance\@colroom-\ht\LT@foot
      \dimen@\pagegoal\advance\dimen@-\ht\LT@foot\pagegoal\dimen@
      \maxdepth\z@
    \fi
  \else
    \advance\vsize-\ht\LT@firstfoot
    \global\advance\@colroom-\ht\LT@firstfoot
    \dimen@\pagegoal\advance\dimen@-\ht\LT@firstfoot\pagegoal\dimen@
    \maxdepth\z@
  \fi
  \ifvoid\LT@firsthead\copy\LT@head\else\box\LT@firsthead\fi\nobreak
  \output{\LT@output}%
}
\def\LT@output{%
  \ifnum\outputpenalty <-\@Mi
    \ifnum\outputpenalty > -\LT@end@pen
      \LT@err{floats and marginpars not allowed in a longtable}\@ehc
    \else
      \setbox\z@\vbox{\unvbox\@cclv}%
      \ifdim \ht\LT@lastfoot>\ht\LT@foot
        \dimen@\pagegoal
        \advance\dimen@-\ht\LT@lastfoot
        \ifdim\dimen@<\ht\z@
          \setbox\@cclv\vbox{\unvbox\z@\copy\LT@foot\vss}%
          \@makecol
          \@outputpage
          \setbox\z@\vbox{\box\LT@head}%
        \fi
      \fi  
      \global\@colroom\@colht
      \global\vsize\@colht   
      \vbox
        {\unvbox\z@\box\ifvoid\LT@lastfoot\LT@foot\else\LT@lastfoot\fi}%
    \fi
  \else
    \ifvoid\LT@firstfoot
      \setbox\@cclv\vbox{\unvbox\@cclv\copy\LT@foot\vss}%
      \@makecol
      \@outputpage
      \global\vsize\@colroom
    \else
      \setbox\@cclv\vbox{\unvbox\@cclv\box\LT@firstfoot\vss}%
      \@makecol
      \@outputpage
      \global\advance\@colroom\LT@footdiff
      \global\vsize\@colroom
    \fi
    \copy\LT@head\nobreak
  \fi
}
\makeatother

\begin{document}

   \title{Tango of celestial dancers: A sample of detached eclipsing binary systems containing $g$-mode pulsating components}
   \subtitle{A case study of KIC9850387}
   \titlerunning{A sample of pulsating eclipsing binaries exhibiting $g$-modes and the case study of KIC9850387}


\author{
    S.~Sekaran\inst{\ref{ivs}}
    \and
    A.~Tkachenko\inst{\ref{ivs}}
    \and
    M.~Abdul-Masih\inst{\ref{ivs}}
    \and
    A.~Pr\v sa\inst{\ref{villanova}}
    \and
    C.~Johnston\inst{\ref{ivs}}
    \and
    D.~Huber\inst{\ref{hawaii}}
    \and
    S.~J.~Murphy\inst{\ref{Sydney},\ref{SAC}}
    \and
    G.~Banyard\inst{\ref{ivs}}
    \and
    A. W. Howard\inst{\ref{CIT}}
    \and
    H. Isaacson\inst{\ref{UCB},\ref{USQ}}
    \and
    D.~M.~Bowman\inst{\ref{ivs}}
    \and
    C.~Aerts\inst{\ref{ivs}, \ref{conny}, \ref{MPI}}
}


\institute{
    Instituut voor Sterrenkunde (IvS), KU Leuven, Celestijnenlaan 200D, B-3001 Leuven, Belgium \\
    \email{sanjay.sekaran@kuleuven.be}
    \label{ivs}
    \and
    Villanova University, Dept.~of Astrophysics and Planetary Science, 800 Lancaster Ave, Villanova PA 19085
    \label{villanova}
    \and
    Institute for Astronomy, University of Hawai‘i, 2680 Woodlawn Drive, Honolulu, HI 96822, USA
    \label{hawaii}
    \and
    Sydney Institute for Astronomy (SIfA), School of Physics, University of Sydney, NSW 2006, Australia
    \label{Sydney}
    \and
    Stellar Astrophysics Centre, Department of Physics and Astronomy, Aarhus University, 8000 Aarhus C, Denmark
    \label{SAC}
    \and
    Department of Astronomy, California Institute of Technology, Pasadena, CA 91125, USA
    \label{CIT}
    \and
    Department of Astronomy,  University of California Berkeley, Berkeley CA 94720, USA
    \label{UCB}
    \and
    Centre for Astrophysics, University of Southern Queensland, Toowoomba, QLD, Australia
    \label{USQ}
    \and
    Department of Astrophysics, IMAPP, Radboud University Nijmegen, NL-6500 GL, Nijmegen, the Netherlands
    \label{conny}
    \and
    Max Planck Institute for Astronomy, Koenigstuhl 17, 69117 Heidelberg, Germany
    \label{MPI}
    }

   \date{Received July 21, 2020; accepted September 28, 2020}

 
  \abstract
   {Eclipsing binary systems with components that pulsate in gravity modes ($g$ modes) allow for simultaneous and independent constraints of the chemical mixing profiles of stars. The high precision of the dynamical masses and radii as well as the imposition of identical initial chemical compositions and equivalent ages provide strong constraints during the modelling of $g$-mode period-spacing patterns.}
   {We aim to assemble a sample of $g$-mode pulsators in detached eclipsing binaries with the purpose of finding good candidates for future evolutionary and asteroseismic modelling. In addition, we present a case study of the eclipsing binary KIC9850387, identified as our most promising candidate, and detail the results of the observational spectroscopic, photometric, and asteroseismic analysis of the system.}
   {We selected all of the detached eclipsing binaries in the \textit{Kepler} eclipsing binary catalogue with \textit{Kepler} Input Catalogue (KIC) temperatures between 6000~K and 10000~K, and performed a visual inspection to determine the presence and density of $g$ modes, and the presence of $g$-mode period-spacing patterns in their frequency spectra. We then characterised our sample based on their $g$-mode pulsational parameters and binary and atmospheric parameters. A spectroscopic follow-up of our most promising candidate was then performed, and the orbital elements of the system were extracted. We then performed spectral disentangling followed by atmospheric modelling and abundance analysis for the primary star. We utilised an iterative approach to simultaneously optimise the pulsational and eclipse models, and subsequently performed an analysis of the pressure- ($p$-) and $g$-mode pulsational frequencies.}
   {We compiled a sample of 93 \textit{Kepler} eclipsing binary stars with $g$-mode pulsating components and identified clear $g$-mode period-spacing patterns in the frequency spectra of seven of these systems. We also identified 11 systems that contained hybrid $p$- and $g$-mode pulsators. We found that the $g$-mode pulsational parameters and the binary and atmospheric parameters of our sample are weakly correlated at best, as expected for detached main-sequence binaries. We find that the eclipsing binary KIC9850387 is a double-lined spectroscopic binary in a near-circular orbit with a hybrid $p$- and $g$-mode pulsating primary with $M_{\text{p}}=1.66_{-0.01}^{+0.01}$ $M_{\odot}$ and $R_{\text{p}}=2.154_{-0.004}^{+0.002}$ $R_{\odot}$, and a solar-like secondary with $M_{\text{s}}=1.062_{-0.005}^{+0.003}$ $M_{\odot}$ and $R_{\text{s}}=1.081_{-0.002}^{+0.003}$ $R_{\odot}$. We find $\ell=1$ and $\ell=2$ period-spacing patterns in the frequency spectrum of KIC9850387 spanning more than ten radial orders each, which will allow for stringent constraints of stellar structure during future asteroseismic modelling.}
   {}

   \keywords{stars: individual: KIC9850387 -- binaries: eclipsing -- binaries: spectroscopic -- stars: fundamental parameters -- stars: oscillations -- asteroseismology}

   \maketitle

\section{Introduction}
\label{sec: intro}

The reputation of eclipsing binaries in providing the "royal road to success" \citep{Russell1948} in stellar astrophysics is well-deserved: The combined analysis of the timeseries of spectroscopic and photometric data enables the determination of the masses and radii of the individual components to a precision of 1\% or better \citep{Torres2010}. These so-called dynamical parameters add to the already powerful prescriptions of identical initial chemical composition and equivalent ages of the individual components provided by binarity, the combination of which provides strong constraints for the calibration of stellar structural and evolutionary models (e.g. \citealt{Ribas2000,Torres2010,Torres2014,Tkachenko2014a,Tkachenko2014b,Stancliffe2015,Claret2018,Constantino2018,Johnston2019a,Johnston2019b,Tkachenko2020}). 

One of the principal aspects of the evolutionary models that is being calibrated in these studies (at present) is the morphology of the chemical mixing profiles within the stellar structures, particularly in the boundary regions of stars with convective cores (e.g. \citealt{Pols1995,Schneider2014,Higl2017,Tkachenko2020}). It has been postulated that the longstanding binary mass-discrepancy problem, the discrepancy between dynamical masses and those derived from evolutionary models (first presented by \citealt{Ribas2000}), is a result of insufficient core-boundary mixing in the evolutionary models: Studies such as \cite{Higl2017} and \cite{Tkachenko2020} have shown that the inclusion of a properly calibrated core-boundary mixing profile in the evolutionary models significantly decreases the observed magnitudes of binary mass discrepancy for stars that have convective cores.

Another method through which one can calibrate internal mixing profiles is asteroseismology \citep{Aerts2010}, particularly the study of low-frequency gravity-mode ($g$-mode) pulsations (e.g. \citealt{Pedersen2018,Michielsen2019}). These modes are typically exhibited by intermediate-mass main-sequence stars above $\sim1.2$ M$_{\odot}$ that are born with convective cores and are excited by either the flux-blocking mechanism at the base of the small ($<10\%$ of the stellar radius) convective region of the outer envelope \citep{Guzik2000, Dupret2005} or the $\kappa$-mechanism \citep{Dziembowski1993} for stars with purely radiative envelopes. Due to the largely radiative (and therefore stably stratified) nature of the envelopes of intermediate-mass stars, these $g$ modes are able to propagate all the way from the near-core region to the surface, unlike in lower-mass stars (cf. Figures 1.7 and 1.8 of \citealt{Aerts2010}). It is this propensity that makes $g$-mode pulsations particularly sensitive to near-core mixing phenomena and chemical stratification \citep{Miglio2008}.

It was theoretically predicted by \cite{Tassoul1980} that in non-rotating, chemically homogeneous stars with a convective core and a radiative envelope, $g$ modes of high radial order ($n>>\ell$, where $\ell$ is the spherical harmonic degree of the mode) are equally spaced in period. The expression for this so-called asymptotic period spacing is: 

\begin{equation}
\hfill \Pi_{\ell} = \frac{\Pi_{0}}{\sqrt{\ell(\ell+1)}},\hfill
\label{eq: Piell}
\end{equation}

\noindent where,

\begin{equation}
\hfill \Pi_{0} = 2\pi^{2}\left(\int_{r_{1}}^{r_{2}}N\frac{dr}{r}\right)^{-1}.\hfill
\label{eq: Pinaught}
\end{equation}

\noindent In these equations, $r$ is the distance from the stellar centre, $N$ is the Brunt-V{\"a}is{\"a}l{\"a} frequency and $r_{1}$ and $r_{2}$ are the radial boundaries of the $g$-mode propagation cavity in the star. These equations demonstrate that the asymptotic period spacing is indeed sensitive to the local conditions in the regions in which these modes propagate.  
\cite{Miglio2008} further expanded upon the theoretical predictions of \cite{Tassoul1980}, showing that periodic dips (i.e. intermittent decreases in the period spacing between modes of consecutive radial orders) in the pattern occur when chemical gradients are present in the stellar interior, with the radial location of the gradient affecting the periodicity and the magnitude of the gradient affecting the magnitude of each dip. \cite{Bouabid2013} then included the effects of diffusive mixing and rotation on the period-spacing pattern using the framework of the traditional approximation of rotation \citep{Townsend2003a,Townsend2003b}. Their conclusions were that 1) mixing reduces the steepness of the chemical gradients in the stellar interior, and therefore reduces the depth of the dips in the $g$-mode period spacing pattern; and that 2) rotation introduces a slope into the pattern by shifting the periods based on the azimuthal order ($m$) of the mode.

The full interior-probing potential of $g$-mode period-spacing patterns was only unlocked after the advent of high-precision, high duty cycle, space-based photometric data such as those provided by CoRoT \citep{Auvergne2009}, \textit{Kepler} \citep{Borucki2010}, K2 \citep{Howell2014}, BRITE \citep{Weiss2014} and TESS \citep{Ricker2015}. These data do not suffer from the aliasing and low duty cycle of ground-based data, enabling the unambiguous identification and characterisation of pulsational frequencies extracted from the photometric data. Following the first detection of period-spacing patterns in the CoRoT photometry by \cite{Degroote2010}, a whole host of studies involving the detection and modelling of period-spacing patterns have been published, from individual case studies (e.g. \citealt{Chapellier2012,Papics2012b,Papics2014,Papics2015,Kurtz2014,Saio2015,Murphy2016,KW2017,Zwintz2017}) to ensembles of a handful to hundreds of stars (e.g. \citealt{Bedding2015,VanReeth2015a,VanReeth2015b,VanReeth2016,Ouazzani2017,Papics2017,Mombarg2019,Li2019a,Li2019b,Li2020a}). These studies reveal a large range of observed radial orders for dipole $g$ modes, covering $n_{\text{g}}\in[10,100]$ (see \citealt{Aerts2020} for a review).

One of the weaknesses of the period-spacing pattern as a diagnostic is the degeneracy between the free parameters (e.g. mass, age, metallicity, chemical composition, interior mixing profiles) used in the evolutionary models (e.g. \citealt{Valle2017}) from which the theoretical patterns are derived (e.g. \citealt{Moravveji2015}). This means that evolutionary models with different input parameters may result in very similar theoretical period-spacing patterns. While it is possible to alleviate some of these degeneracies through a proper statistical treatment (e.g. \citealt{Aerts2018}), ideally the inclusion of independent constraints on the various free parameters should be considered. In the single-star case, this can take the form of spectroscopic constraints on the effective temperature, surface gravity and metallicity, as well as luminosity constraints from the astrometric data provided by the \textit{Gaia} space mission (\citealt{Gaia2016}, see \citealt{Pedersen2020} for details on the proper treatment of Gaia luminosities). However, as mentioned, far more stringent constraints are afforded by binarity, specifically in the form of the highly precise dynamical parameters that can be extracted from detached eclipsing binaries with main-sequence components. \cite{Johnston2019b} have shown how the inclusion of binary information significantly improves the constraining of stellar models when combined with asteroseismic information.

The complementary nature of binarity and $g$-mode asteroseismology is well-known, with a number of individual case studies on particular stars (e.g. \citealt{Maceroni2009,Maceroni2013,Welsh2011,Chapellier2013,Debosscher2013,Hambleton2013,Keen2015,Schmid2015,Schmid2016,Matson2016,Lee2018,Zhang2018,Guo2019b,Guo2017a,Guo2017c,Guo2019a,Guo2020, Zhang2020}). However, these case studies tend to be either 1) purely observational, 2) feature only tidally excited $g$ modes (e.g. \citealt{Guo2017a}), 3) involve non-eclipsing binaries (e.g. \citealt{Schmid2015}), or 4) do not report the detection of period-spacing patterns (e.g. \citealt{Debosscher2013}). Also, of these studies only \cite{Schmid2016}, \cite{Zhang2018}, \cite{Guo2019b} and \cite{Zhang2020} include detailed asteroseismic modelling, with \cite{Schmid2016} presenting the most detailed modelling effort of the three as they used theoretical period-spacing patterns, derived from evolutionary models, to match their observed period-spacing patterns. This provides stronger constraints on the stellar structure when compared to, for example, fitting for $\Pi_{0}$ \citep{Ouazzani2019,Mombarg2020}.  

Overall, there is a lack of studies that combine detailed asteroseismic modelling with eclipsing binary analysis. This would enable two independent calibrations of the amount of core-boundary mixing that would have to be included in the evolutionary models in order to match either the dynamical parameters or the observed $g$-mode period-spacing patterns, enabling simultaneous confrontation of dynamical, evolutionary and asteroseismic parameters. In addition, there has thus far only been one ensemble study of eclipsing binaries with $g$-mode period-spacing patterns \citep{Li2020a}, which is a rather curious phenomenon considering that hundreds of stars with $g$-mode period-spacing patterns have been discovered \citep{Li2020b}.  

\cite{Gaulme2019} performed a systematic search for pulsators in \textit{Kepler} eclipsing binary systems, reporting a total of 115 $g$-mode pulsators ($\gamma$ Doradus). They did not investigate the asteroseismic viability of said stars or perform any sort of asteroseismic or statistical analysis of their $g$-mode pulsator sample. In this paper, we present a sample of detached eclipsing binaries with excellent $g$-mode asteroseismic potential, which we define as the detection of $g$-mode period-spacing patterns that span six radial orders or more without any gaps in the pattern, as short period-spacing patterns and those with gaps result in additional degeneracy during asteroseismic modelling. These stars are identified through an independent systematic search of the \textit{Kepler} Eclipsing Binary Catalogue (KEBC, \citealt{Prsa2011,Slawson2011,Kirk2016,AbdulMasih2016}). 

In addition, we present the observational spectroscopic, photometric and asteroseismic analysis of the pulsating eclipsing binary KIC9850387 as a case study of the most promising candidate for future evolutionary and asteroseismic modelling. Section \ref{sec: targets} details the sample selection process and an introduction to KIC9850387, Section \ref{sec: spectroscopy} details our spectroscopic follow-up and analysis, and Section \ref{sec: photometry} details the photometric analysis of the system. The asteroseismic analysis of the hosted pulsating star is presented in Section \ref{sec: pulsations} and we present a discussion of our results and conclusions in Section \ref{sec: conclusions}.

\section{Identifying optimal targets}
\label{sec: targets}

To identify optimal targets for asteroseismic analysis, we first selected all eclipsing binaries observed by \textit{Kepler} during its nominal mission in the KEBC. We selected all stars with Kepler Input Catalogue (KIC, \citealt{Brown2011}) effective temperatures between 6000 and 10000 K that had morphology parameter values below 0.5. The morphology parameter in the KEBC indicates the degree of 'detachedness' of an eclipsing system \citep{Matijevic2012}. This parameter is calculated by an automated classification algorithm and takes values between 0 and 1, with overcontact systems being assigned a value of 1 and completely detached systems being assigned a value of 0. \cite{Matijevic2012} had compared their morphology parameter results with a manual classification and concluded that those systems that scored below 0.5 can be considered to be predominantly detached, and as such we restricted our analysis to those systems to ensure that any pulsational signature detected could be disentangled from binary evolutionary effects such as mass transfer (e.g. \citealt{Niemczura2017}). 

\begin{figure*}[!t]
\begin{center}
\includegraphics{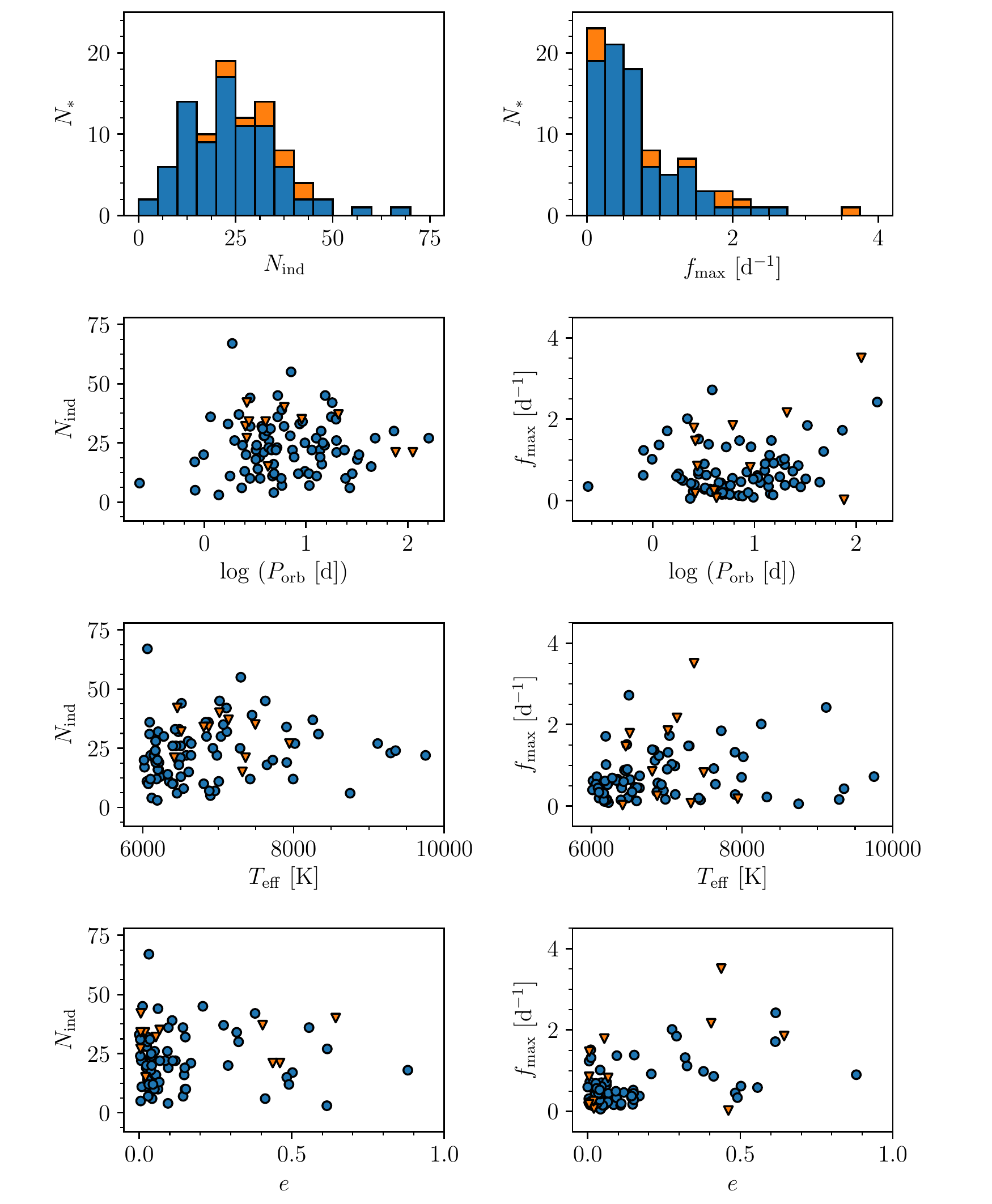}
\caption[Pulsational versus binary/atmospheric properties of the eclipsing binaries with $g$-modes.]{Pulsational versus binary/atmospheric properties of the eclipsing binaries with $g$-modes. The plots in the left column show the distributions of $N_{\text{ind}}$ and those in the right column show the distributions of $f_{\text{max}}$ . The first row is a display of stacked histograms of $N_{\text{ind}}$ and $f_{\text{max}}$, with the vertical axis representing the number of systems ($N_{*}$). For the first row of plots, systems where only $g$ modes were detected are represented in blue, and systems where both $p$ and $g$ modes were detected are represented in orange. The 2nd, 3rd and 4th rows are displays of the variation of $f_{\text{max}}$ and $N_{\text{ind}}$ with respect to log $P_{\text{orb}}$ (2nd row), $T_{\text{eff}}$ (3rd row) and $e$ (4th row). For the 2nd, 3rd and 4th rows of plots, stars where only $g$ modes were detected are represented by blue circles, and stars where both $p$ and $g$ modes were detected are represented by orange triangles.}
\label{fig: EB_Sample}
\end{center}
\end{figure*}

\begin{figure}[ht!]
\includegraphics[width=\hsize]{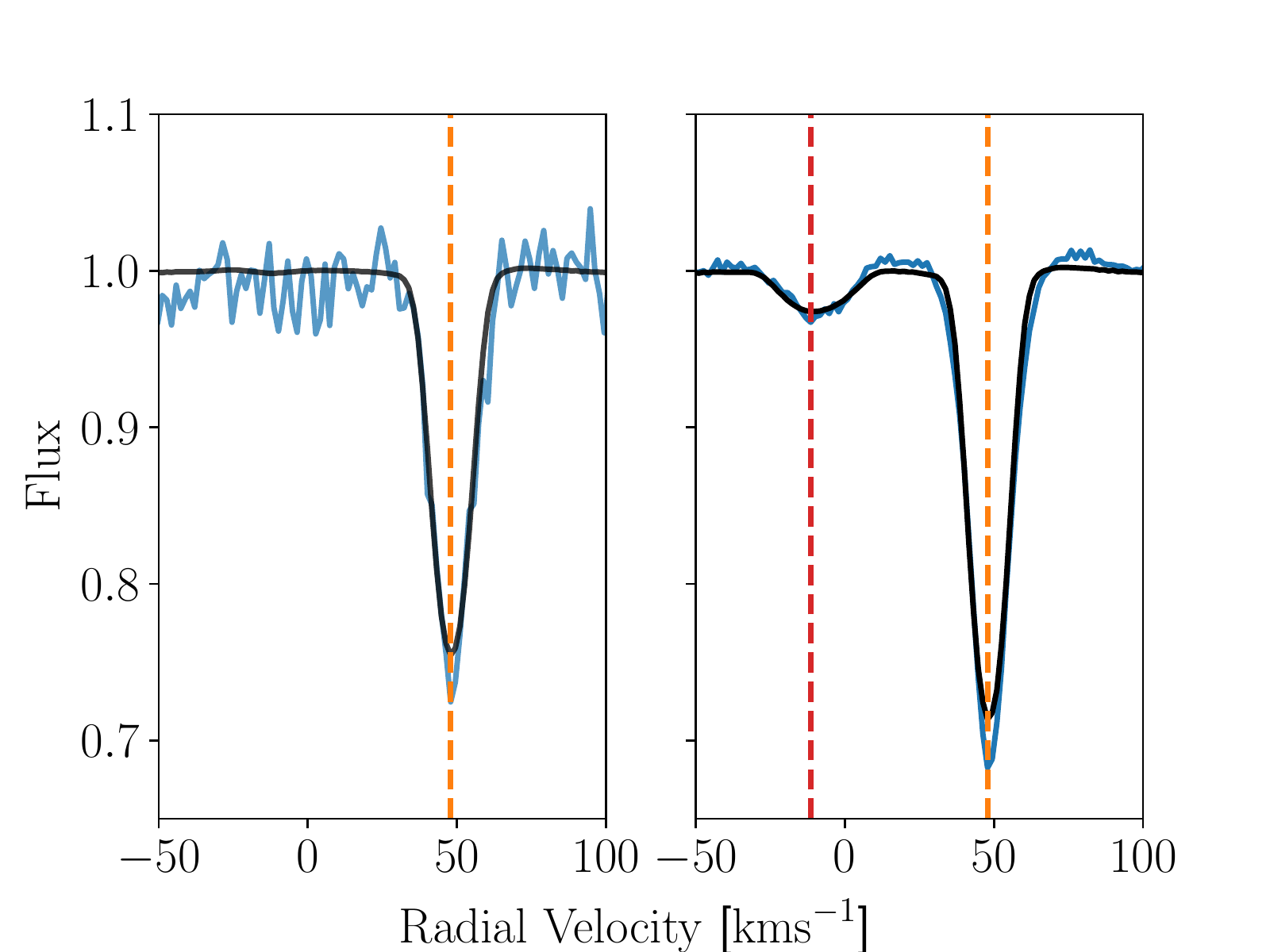}
\caption[Comparison of HERMES and HIRES LSD profiles at an orbital phase of $\sim -0.45$.]{Comparison of the LSD profiles generated from a HERMES (left panel) and a HIRES (right panel) spectrum of KIC9850387 taken at an orbital phase of $\sim -0.45$. Synthetic LSD profiles in black are fitted to the observed LSD profiles in blue in order to determine the radial velocities of the primary (indicated by the vertical orange dashed lines) and the secondary (indicated by the vertical red dashed line) component.}
\label{fig: HermesVsHires}
\end{figure}

We then performed a pulsational screening of the sample using the following steps: 1) Prewhiten the long-cadence light curves with the first 1000 orbital harmonics, using the detrended simple aperture photometry (SAP) fluxes and the orbital frequencies provided by the KEBC; 2) Manually inspect the residual periodograms for high-amplitude low-frequency peaks that correspond to $g$-mode pulsations; 3) Visually identify periodograms that contain dense clusters (six or more) of $g$-mode frequency peaks, which increases the likelihood of finding period-spacing patterns; 4) Perform a manual period-spacing search in the orbital-harmonic-removed frequency spectra of the best candidates.

\subsection{Sample characterisation}
\label{subsec: EB_sample}

Based on our selection criteria, we performed pulsational screening on a total of 296 eclipsing binaries listed in the KEBC. After step 2 of our pulsational screening, we identified $g$-mode pulsations in 93 of those systems. Of these systems, 24 exhibited dense clusters of $g$-mode frequencies in the Fourier domain. Eleven out of these 24 systems also showed prominent pressure ($p$) modes. After performing the period-spacing search (step 4), we identified candidate period-spacing patterns in seven out of the 24 systems, with only two displaying continuous patterns of longer than six radial orders. We have therefore assembled a sample of pulsators in eclipsing binary systems displaying a variety of pulsational attributes that are of asteroseismic interest, but whose analysis is outside the scope of this paper. Our goal was to identify targets that were clearly strong in terms of $g$-mode asteroseismic potential, and based on that criterion, we selected only the two out of the 296 targets displaying period-spacing patterns that fulfilled our criteria.

In addition to the pulsational screening, we also characterised the sub-sample of 93 eclipsing binary systems with $g$-mode pulsating components in terms of its pulsational properties and binary and atmospheric properties, to determine if there were any correlations between the two types of properties. The pulsational characterisation we performed comprised: 1) determining the frequency of highest amplitude ($f_{\text{max}}$) in the $g$-mode regime (defined to be the region below 5 d$^{-1}$), and 2) calculating the number of independent frequencies ($N_{\text{ind}}$) in said $g$-mode regime (see Section \ref{sec: pulsations} for a detailed description our methodology for the identification of independent frequencies). 

At present, the KEBC does not provide additional binary orbital information outside of the orbital period ($P_{\text{orb}}$), morphology (Morph), and the timestamp of the first superior conjunction. However, one can derive estimates of the eccentricity ($e$) and the argument of periastron ($\omega$) directly from the photometry itself, as $e \cos \omega$ is proportional to the phase separation of the eclipses ($\Delta\phi$) and $e \sin \omega$ is proportional to the ratio of the primary ($p$) and secondary ($s$) eclipse widths according to $(w_{\text{p}} - w_{\text{s}})/(w_{\text{p}} + w_{\text{s}})$ in the phase domain (see e.g. \citealt{Prsa2018} for more information). Pr{\v{s}}a et al. (in prep.) devised an approximation of the morphology of an eclipse through a combination of linear and quadratic functions, and fitted this functional combination to the eclipse regions of all doubly eclipsing (i.e. displaying both a primary and a secondary eclipse) systems in the KEBC with sufficiently high signal-to-noise ratios (S/N) to determine the ingresses, egresses and midpoints of each eclipse. This provided the inputs to obtain $\Delta\phi$ and $(w_{\text{p}} - w_{\text{s}})/(w_{\text{p}} + w_{\text{s}})$, and therefore $e$ and $\omega$. We were therefore able to obtain $e$ and $\omega$ for 81 out of 93 of the selected eclipsing binary systems with $g$-mode pulsating components that fulfilled the criteria of Pr{\v{s}}a et al. A full description of the methodology will be presented in Pr{\v{s}}a et al. (in prep.). 

Table \ref{tab: sample} displays the KIC IDs, pulsational ($N_{\text{ind}}$ and $f_{\text{max}}$) and binary/atmospheric (log $P_{\text{orb}}$, Morph, $T_{\text{eff}}$, $e$, and $\omega$) parameters\footnote{log $P_{\text{orb}}$, Morph, $T_{\text{eff}}$ are taken from the KEBC, and $e$ and $\omega$ are determined by Pr{\v{s}}a et al. (in prep.).} of the 93 eclipsing binary systems with $g$-mode pulsating components in our sample, as well as whether $p$-mode frequencies were also observed in their frequency spectra. We also include a selection of eight phase-folded light curves of the systems in our sample in Figure \ref{fig: LC_selection} to showcase the variety of morphologies exhibited by the light curves of our sample. 32 of the stars identified during our analysis have been previously reported in the study of \cite{Gaulme2019}. In addition, 13 other systems, not listed in \cite{Gaulme2019}, have also been previously studied in an asteroseismic context. For these cases, we also include in Table \ref{tab: sample} all of the references to the studies that refer to the discovery or analysis of $g$ modes. The 45 systems without a reference in Table \ref{tab: sample} are therefore eclipsing binary systems whose $g$-mode pulsational characteristics were discovered during this study.

\cite{Li2020a} performed a study of 35 \textit{Kepler} eclipsing and spectroscopic binaries in which they identified clear period-spacing patterns, with the majority of those selected from the catalogue of \cite{Gaulme2019}. 11 of those are also in our sample as part of the 24 systems whose periodograms contain dense clusters of $g$-mode frequencies. Of these 11 systems, we only found period-spacing patterns in seven of them (as mentioned earlier in this section). We suspect that the reason that they found more period-spacing patterns than we did is due to a difference in methodology: Li et al had used a S/N cutoff of 3.5 while we used the more-conservative $\text{S/N}=4$ \citep{Breger1993}, and they subtracted binned and phase-folded light curves from the original timeseries, while we prewhitened the orbital harmonics to remove the binary signature. Our approach would necessarily result in a lower number of extracted frequencies per star (due to the higher S/N cutoff) and a potential reduction in pulsational amplitude and therefore S/N (due to the orbital-harmonic prewhitening) when compared to the approach of Li et al. 

The stars that are in their sample, but not ours, have one or more of the following characteristics: 1) Non-eclipsing; 2) Not clearly detached ($\text{Morph}>0.5$); 3) Undefined morphology (i.e. $\text{Morph}=-1$); 4) Undefined KIC temperatures ($T_{\text{eff}}=-1$); 5) Higher-order eclipsing system (the quintuple system KIC4150611). We posit that our catalogue and that of \cite{Li2020a} are complementary, as we include all $g$-mode (and hybrid $p$- and $g$-mode) pulsators independently identified in detached eclipsing binaries, with both pulsational and orbital characterisation. 

Figure \ref{fig: EB_Sample} shows various distributions of pulsational parameters ($N_{\text{ind}}$ and $f_{\text{max}}$) versus the binary and atmospheric (log $P_{\text{orb}}$, $T_{\text{eff}}$, $e$) parameters. We did not include the distributions with respect to Morph and $\omega$ as our sample consists only of detached binaries. Given that $\omega$ is simply a spatial orientation parameter, any potential correlation would simply be pure coincidence. We also calculated the Spearman's rank correlation ($\rho$) to test if there were any correlations between the pulsational parameters and binary and atmospheric parameters (other than $\omega$). We obtained $|\rho|$ values below 0.3 for all combinations of pulsational parameters and binary and atmospheric parameters parameters, with $p$-values of the null hypothesis (that the parameters are uncorrelated) larger than 0.1 except for $f_{\text{max}}$/$T_{\text{eff}}$, $N_{\text{ind}}$/$T_{\text{eff}}$ and $f_{\text{max}}$/$e$, where the $p$-values are larger than 0.01. These values indicate that at best, there is only a weak correlation between any of the $g$-mode pulsational parameters and binary and atmospheric parameters, as expected for detached main-sequence binaries.

\subsection{KIC9850387: An eclipsing binary with multimodal period-spacing patterns}
\label{subsec: KIC9850387}

Of the two systems that we considered to be our best candidates, KIC9850387 was very obviously the better one: Our preliminary analysis revealed two $g$-mode period-spacing patterns of more than eight radial orders each (the longest in our sample) corresponding to different $\ell$ values in its frequency spectrum, as well as a few high-amplitude $p$ modes. According to the KEBC, KIC9850387 has a period of 2.74 d and a morphology value of 0.47 with evenly spaced eclipses of near-equal widths, indicative of a circular or near-circular orbit. The KIC parameters for this system are as follows: $K_{\text{mag}} = 13.549$, $T_{\text{eff}} = 6808$ K, log $g$ = 4.028, $R$ = 1.818 R$_{\odot}$, and $\text{[Fe/H]} = -0.291$ \citep{Brown2011}. It should be noted that since these parameters were derived from photometric colours, they are likely inaccurate in general, and are probably even worse for binary systems. However, we list them here for completeness. KIC9850387 was first studied by \cite{Zhang2020}, who claimed that the system was a pre-main-sequence SB1 eclipsing system containing a hybrid $p$- and $g$-mode pulsator, and reported the detection of an $\ell=1$ period-spacing pattern with a mean period spacing of $2756.2\pm0.8$~s. This system was also studied by \cite{Li2020a}, who had reported a $\ell=1$, $m=1$ and a $\ell=2$, $m=2$ period-spacing pattern, with $\Pi_{0}=3894\pm7$~s and a core-rotation rate $f_{\text{rot,core}}=0.0053$~d$^{-1}$.

\section{High-resolution spectroscopy}
\label{sec: spectroscopy}

We embarked on a dedicated spectroscopic follow-up campaign of KIC9850387 using the HERMES spectrograph \citep{Raskin2011} attached to the 1.2-m Mercator telescope at the Roque de los Muchachos observatory on La Palma, Spain. Due to the relative faintness of the star ($K_{\text{mag}} = 13.549$) and the 1.2-m diameter of the Mercator, it was difficult to obtain spectra with a S/N above 20, with short-enough exposure times to prevent excessive line smearing due to the low orbital period ($P=2.74$ d). 

However, through the use of techniques such as least-squares deconvolution (LSD, as described in \citealt{Tkachenko2013b}) and spectral disentangling (see Section \ref{subsec: SPD}), we were confident of being able to, at the very least, obtain precise radial velocities from the spectra, if not atmospheric parameters. The principle assumption of the LSD technique is that the observed spectrum is a convolution of a mean line profile with a predetermined line mask \citep{Donati1997}, which is a template of delta functions with wavelengths and line depths corresponding to a synthetic spectrum. LSD is the solution of the \textit{inverse} problem, which is the determination of the mean line profile given an observed spectrum and a line mask. 

This LSD profile has a significantly increased S/N when compared to any single spectral line, scaling with the square root of the number of spectral lines used in the mask that are also in the observed spectrum, therefore enabling more precise determinations of radial velocities. Tests with LSD profiles determined from synthetic spectra with white noise added to resemble spectra with $\mathrm{S/N}=20$ resulted in radial velocity precisions of the order of 0.4~${\text{km\ s}^{-1}}$. These precisions are acceptable for stars with low orbital periods because they tend to have high radial velocity semi-amplitudes of the order of 100 km~s$^{-1}$.

We obtained a total of 18 spectra between April and September 2019 with the HERMES spectrograph, and generated LSD profiles from the normalised spectra using the line mask of a dwarf star with $T_{\text{eff}}=7000$ K, which is reasonably close to the KIC $T_{\text{eff}}=6808$ K. This line mask comprises more than 3000 lines, and was calculated using the \textsc{gssp} software package (described in Section \ref{subsec: atm}). Unfortunately, we were only able to clearly visually discern the signature of the primary in each profile. The eclipse depth ratios indicated that the secondary should be a G-type dwarf, assuming that the primary is an F-type dwarf as indicated by the KIC $T_{\text{eff}}$ and log $g$. As such it would be much fainter than the primary and therefore have a low light contribution, which when combined with the low S/N, results in the signature of the secondary being indistinguishable from the noise.

Since \cite{Zhang2020} claimed that the system was a SB1, based on only six low-resolution ($R\sim1800$) LAMOST spectra \citep{DeCat2015}, we decided to check their claims. To that end, we obtained between October and November 2019 an additional eight spectra with a $\mathrm{S/N}\sim50$ from the HIRES spectrograph \citep{Vogt1994} attached to the 10-m Keck Telescope at the Mauna Kea Observatories in Hawai‘i, USA. Once again, we calculated the LSD profiles for these new spectra, and signatures of both components were visually discernible, although those of the secondary were clearly less distinct (see Figure \ref{fig: HermesVsHires} for a comparison of the HERMES and HIRES LSD profiles at similar orbital phases). It is therefore conclusive that KIC9850387 is an eclipsing, double-lined spectroscopic binary (SB2) system.


\subsection{Spectroscopic orbital elements}
\label{subsec: orbital}

We obtained the radial velocities for each component by fitting synthetic LSD profiles to the observed HERMES and HIRES LSD profiles. We fitted the 18 HERMES LSD profiles as if the system was a SB1 (as we are unable to discern the secondary) and the eight HIRES LSD profiles as a SB2. We used a grid of synthetic LSD profiles to fit our observed profiles, calculated from synthetic spectra with $T_{\text{eff}}=7000$ K, log $g=4.0$ dex and [M/H]$=0.0$. We allowed for the $v$ sin $i$ as a proxy of rotational broadening to vary as a free parameter to account for the effects of line-profile variations in the primary star, and included a scaling factor for the depth of each profile to account for the light contribution of each component to the total flux. The best-fitting $v$ sin $i$ values for the primary star ranged between $9-11$ $\text{km\ s}^{-1}$, showing temporal variations of line-broadening consistent with non-radial pulsations \citep{Aerts2014}. In total, we obtained 24 and eight radial velocities for the primary and the secondary components, respectively. Due to the lower light contribution of the secondary, its velocities are less precisely determined than those of the primary.

A preliminary Keplerian orbital fit was then performed using the Markov Chain Monte Carlo (MCMC) routine \texttt{emcee} \citep{ForemanMackey2013} to optimise the orbital elements. We chose to fix the orbital period at the value obtained from the KEBC due to the much longer timebase and the much finer sampling of the \textit{Kepler} data, and similarly, we assumed a circular orbit as indicated by the photometry. This preliminary fit, while not as robust as our final combined fit with both radial velocities and \textit{Kepler} photometry (see Section \ref{subsec: eclipsemodel}), was still relatively good and is a necessary step as it provides constraints for the process of spectral disentangling, particularly the radial velocity semi-amplitudes of the individual components ($K_{1}$ and $K_{2}$). 

\subsection{Spectral disentangling}
\label{subsec: SPD}

The SB2 nature of KIC9850387 requires additional consideration during the determination of atmospheric parameters of the individual components: Each observed spectrum is a sum of the spectra of the individual components that have been 1) Doppler shifted by the radial velocities of the individual components at the time of observation, and 2) scaled by the wavelength-dependent light contribution of the individual components to the total flux, which depends on the spectral energy distribution of the individual components.

One of the ways in which binary stellar spectra can be analysed is by separating them into their individual components, which can be accomplished using the technique known as spectral disentangling \citep{SS1994,Hadrava1995}. Spectral disentangling involves the modelling of the Doppler shift of spectral lines at each orbital phase, enabling the simultaneous determination of both orbital elements and the mean spectrum of each component of the system. This technique has been widely applied to separation of a variety of multiple systems, from single- to multiple-lined systems \citep{Hensberge2000,PH2005,PH2010}. While there are a number of different ways in which spectral disentangling can be performed (see \citealt{PH2010} for a summary of different methodologies), we adopted a Fourier domain-based disentangling procedure implemented in the code FDBinary \citep{Ilijic2004b}.

Due to the possibility of additional systematic effects (such as instrumental wavelength-dependent line-depth variations) on the final result, which are difficult to properly account for when combining the HERMES and HIRES spectra, we performed spectral disentangling on the 18 HERMES spectra and the eight HIRES spectra separately, enabling us to compare and contrast the results from each dataset. We initially attempted to optimise the orbital parameters in the spectral disentangling procedure for both types of spectra. However, we obtained wildly varying values for $K_{2}$ depending on the wavelength region that was being disentangled: This is likely due to the low S/N for the HERMES spectra, and the poor orbital phase coverage for the HIRES spectra, resulting in an inability to properly disentangle the signal of the secondary (which has a low light contribution) from the primary.

As such, we chose to fix the orbital parameters during disentangling at the values obtained from our radial velocity fit. Even though the disentangled component spectra had a significantly higher S/N ($\sim$ 50 for the HERMES and 120 for the HIRES) compared to any single observed spectrum, scaling with square root of the number of observations, only the disentangled primary component spectrum of the each dataset was of sufficiently high S/N for atmospheric parameter determination. The line depths of the disentangled secondary component spectrum of the higher-quality Keck dataset were unfortunately still too low for proper atmospheric parameter determination. However, the eclipsing nature of KIC9850387 means that several secondary atmospheric parameters could be instead determined through the subsequent eclipse modelling process (see Section \ref{subsec: eclipsemodel}), and the disentangled spectrum of the secondary can still be used for a qualitative consistency check for these parameters.

\subsection{Atmospheric parameter determination}
\label{subsec: atm}

\begin{figure}[t]
\includegraphics[width=\hsize]{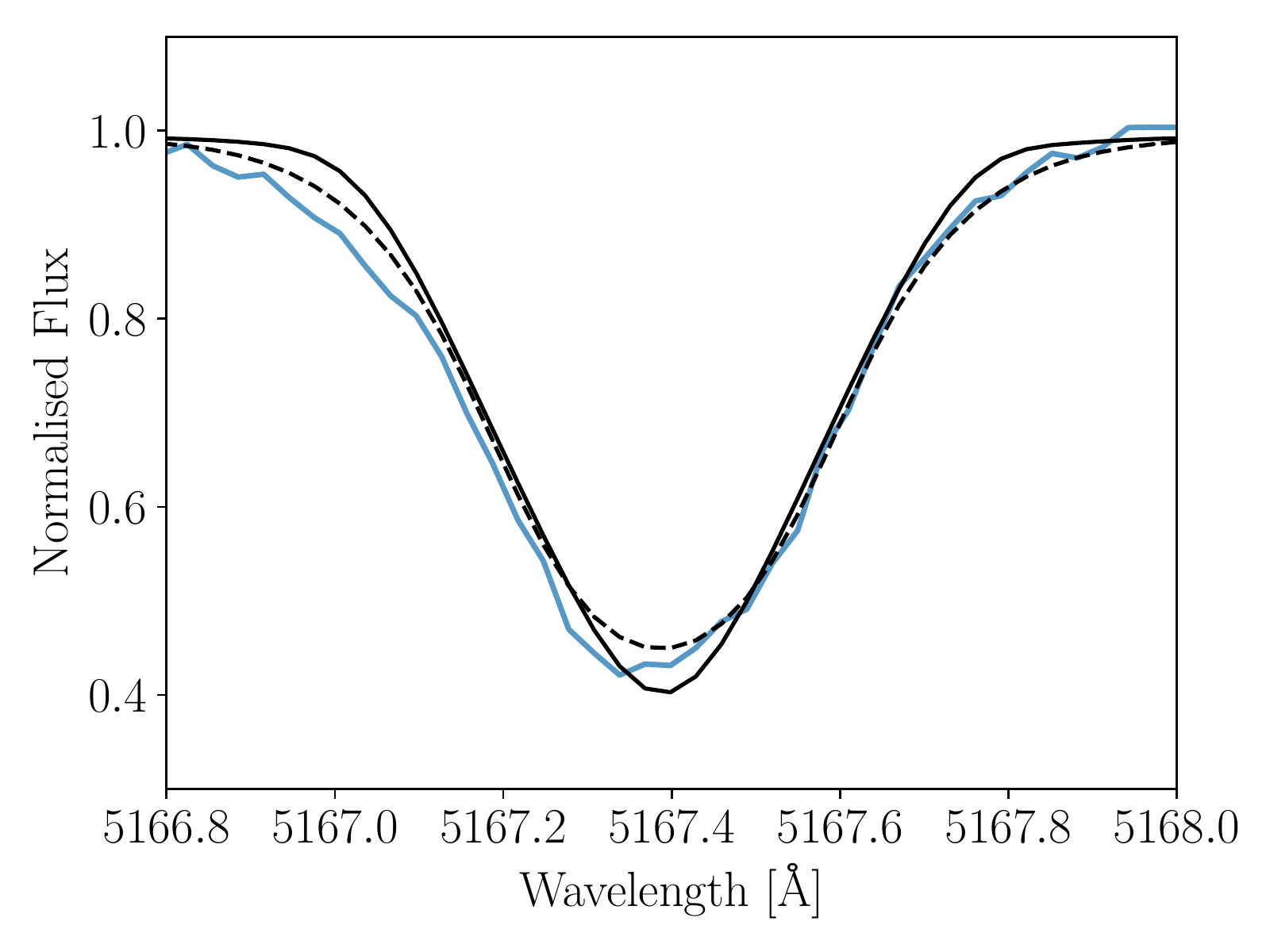}
\caption[Comparison of synthetic spectral fits with and without including $v_{\text{macro}}$.]{Comparison of synthetic spectral fits with (dashed black) and without (solid black) including $v_{\text{macro}}$, compared to a Mg I line of the disentangled primary component HIRES spectrum (in blue).}
\label{fig: vmacro}
\end{figure}

The atmospheric parameters of the primary star were determined by fitting synthetic spectra to our disentangled primary component HERMES and HIRES spectra using the Grid Search in Stellar Parameters (\textsc{gssp}) software package \citep{Tkachenko2015}. As the name suggests, \textsc{gssp} is able to fit a grid of synthetic spectra that are generated on-the-fly using the \textsc{SynthV} radiative transfer code \citep{Tsymbal1996} combined with a grid of atmospheric models from the \textsc{LLmodels} code \citep{Shulyak2004}. The best-fitting values and error estimates of the atmospheric parameters $T_{\text{eff}}$, log $g$, microturbulent velocity ($v_{\text{micro}}$), macroturbulent velocity ($v_{\text{macro}}$), projected rotational velocity ($v_{\text{rot}}$ sin $i$), the global metallicity ($\mathrm{[M/H]}$), and the light contribution of primary component to the total flux $L_{\text{p}}$, can then be determined from the distribution of $\chi^{2}$ values of the fit of each synthetic spectrum to the observed spectrum.

The HERMES spectra that we obtained had significantly higher S/N ($\sim$ 20) in the redder parts of the visual spectrum ($\sim$5000--7000~\AA), and consequently the disentangled primary component spectrum also displayed such behaviour. We therefore chose to fit the disentangled primary component HERMES spectrum in region between 5000 and 6650 {\AA} to minimise the effect of spectral normalisation errors as a result of low S/N, as well as improve the precision of the fit overall. This region also contains the H$_{\alpha}$ line that is essential for the constraining of $T_{\text{eff}}$. To enable a consistent comparison with the HERMES spectra, we also fit the same wavelength region for the HIRES spectra (whose wavelength range unfortunately does not cover the H$_{\beta}$ region). We also chose not to include the H$_{\gamma}$ region as the normalised continuum level drops below unity for AF-type stars in this region, resulting in increased spectral normalisation errors that would inevitably propagate into the disentangled component spectra.

Even though the disentangled primary component HERMES spectrum had significantly improved S/N when compared to the individual observed spectra, it was still too noisy to properly disentangle the various types of velocity broadening (microturbulent, macroturbulent and rotational). Therefore, we chose to fix $v_{\text{micro}}$ at 2.0 km\ s$^{-1}$ and $v_{\text{macro}}$ at 0 km\ s$^{-1}$ (as per \citealt{Tkachenko2013b}), allowing only $v_{\text{rot}}$ sin $i$ to vary in the fit for both the HERMES and HIRES component spectra. This was done in order to obtain a similar basis for the comparison of the HERMES and HIRES results. In addition, $\mathrm{[M/H]}$ and $L_{\text{p}}$ are largely degenerate parameters: $\mathrm{[M/H]}$ affects the basic thermodynamical properties of the star and therefore also affects metal-line depths, and $L_{\text{p}}$ is a global line-depth scaling factor. This degeneracy is further compounded by the fact that log $g$ is not well-constrained in the spectroscopic analysis of F-type stars, as the Balmer lines are largely insensitive to log $g$ in this temperature range. 

To break this degeneracy, constraints would have to be placed on $\mathrm{[M/H]}$ and $L_{\text{p}}$ parameter (and ideally log $g_{\text{p}}$), and since we are able to determine $L_{\text{p}}$ and log $g_{\text{p}}$ with high precision from eclipse modelling, we employ an iterative methodology using both types of spectra to determine the atmospheric parameters: 1) Perform an initial fit in \textsc{gssp} with unconstrained prior ranges for $\mathrm{[M/H]}$ and $L_{\text{p}}$ to obtain an estimate for the primary effective temperature ($T_{\text{eff,p}}$); 2) Perform eclipse modelling (described in Section \ref{subsec: eclipsemodel}) using the $T_{\text{eff,p}}$ estimate to obtain estimates for $L_{\text{p}}$ and log $g$; 3) Fix the $L_{\text{p}}$ and log $g_{\text{p}}$ estimates in \textsc{gssp} and perform another iteration of atmospheric parameter determination to obtain $\mathrm{[M/H]}$ and $T_{\text{eff,p}}$; 4) Iterate between atmospheric analysis and eclipse modelling until the differences in $T_{\text{eff,p}}$, $L_{\text{p}}$ and log $g_{\text{p}}$ between consecutive iterations is less than 1\%.

\begin{table}[b]
\setlength{\tabcolsep}{12pt}
\renewcommand{\arraystretch}{1.3}
\caption[Relative chemical abundances of KIC9850387.]{The relative chemical abundances derived from the disentangled primary component HIRES spectrum of KIC9850387. The first column lists the elements and the second column lists the relative abundance maximum-likelihood estimates and errors based on the 68\% confidence intervals of the fit parameters. The third column lists the solar reference values for the chemical abundances from \cite{Asplund2009}.}
\begin{center}
\begin{tabular}{ ccc }
 \hline
 \hline
 \multicolumn{1}{c}{Element} & \multicolumn{1}{c}{[E/H] (dex)} & \multicolumn{1}{c}{Solar}\\ 
 \hline
 C & $-0.28\pm0.28$ & $-3.57$\\ 
 Mg & $-0.04\pm0.18$ & $-4.40$\\ 
 Si & $-0.31\pm0.30$ & $-4.49$\\ 
 Ca & $0.01\pm0.16$ & $-5.66$\\ 
 Fe & $-0.20\pm0.05$ & $-4.50$\\ 
 Na & $0.28\pm0.46$ & $-5.76$\\ 
 Sc & $0.08\pm0.23$ & $-8.85$\\ 
 Ti & $0.00\pm0.14$ & $-7.05$\\ 
 Cr & $-0.13\pm0.13$ & $-6.36$\\ 
 Y & $0.31\pm0.24$ & $-9.79$\\ 
 Ni & $-0.20\pm0.13$ & $-5.78$\\ 
 \hline
\end{tabular}
\tablefoot{These chemical abundances were derived by fixing the atmospheric parameters at those derived in Table \ref{tab: parameters}, with $v_{\text{macro}}$ set at 0~km~s$^{-1}$.}
\end{center}
\label{tab: abundances}
\end{table}

Using this methodology, we were able to obtain well-constrained spectroscopic determinations of $T_{\text{eff,p}}$, $\mathrm{[M/H]}$ and the projected rotational velocity of the primary component $v_{\text{rot,p}}$ sin $i$, which are the only parameters that we are unable to determine directly from eclipse modelling. It was found within a few iterations that the results were consistent between the HERMES and HIRES component spectra. We therefore decided to perform further iterations only with the higher S/N HIRES spectra, as the results were far more precise. In addition, the different types of velocity broadening were able to be disentangled, and we were able to determine $v_{\text{micro,p}}$ as well. Including $v_{\text{micro,p}}$ as a free parameter as opposed to fixing it at 2.0 km\ s$^{-1}$ had a minimal effect on the other atmospheric parameters: The maximum-likelihood estimates of these parameters were slightly shifted within the error bars, although the error bars themselves were slightly smaller when including $v_{\text{micro,p}}$.

It should be noted that we did not attempt to fit for $v_{\text{macro}}$: It is well known that in F-type stars, $v_{\text{macro}}$ and $v_{\text{rot,p}}$ sin $i$ are degenerate (see \citealt{Fossati2011} for a detailed discussion). Even though this degeneracy is lifted for slow rotators, and a significant improvement in the fit can be obtained upon inclusion of macroturbulence, the poor phase coverage of the HIRES spectra means that a significant amount of pulsational distortion (from the high-amplitude $g$ modes) is present in the disentangled primary component HIRES spectrum\footnote{Spectral disentangling interprets radial velocity variation in the spectral lines as originating completely from orbital motion, and therefore completely ignores pulsational variation.}. This pulsational distortion manifests as asymmetric line-profile variations, which were already noted in the wings of the primary component in the HIRES LSD profiles (see the right panel of Figure \ref{fig: HermesVsHires}), and the inclusion of $v_{\text{macro}}$ in the fit attempts to correct for that. Figure \ref{fig: vmacro} shows the effects of including $v_{\text{macro}}$ in the fit (resulting in the broader wings of the synthetic spectral line with the inclusion of $v_{\text{macro}}$). \cite{Aerts2014} have shown that macroturbulent broadening is able to mimic the effects of pulsational broadening, and that a timeseries of high-resolution spectra would be required to properly disentangle these two types of broadening due to the temporal nature of pulsations. As such, due to the limitations of our dataset, we chose to ignore macroturbulence. 

In addition to the atmospheric parameters of the star, we were also able to determine the abundances for 11 different chemical elements from the disentangled primary component spectrum. \textsc{gssp} can be used to fit for individual chemical abundances once $\mathrm{[M/H]}$ is known. After completing the atmospheric parameter determination, we fixed their values at the maximum-likelihood estimates and varied only the individual chemical abundances. The maximum-likelihood estimates and 68\% confidence intervals of these abundances (relative to the solar values taken from \citealt{Asplund2009}) are listed in Table \ref{tab: abundances}. The relative abundances obtained are largely consistent with $\mathrm{[M/H]}=-0.109$ except for yttrium (Y), which was found to be overabundant by $\sim0.1$ dex.

\begin{figure*}[t]
\includegraphics[width=\hsize]{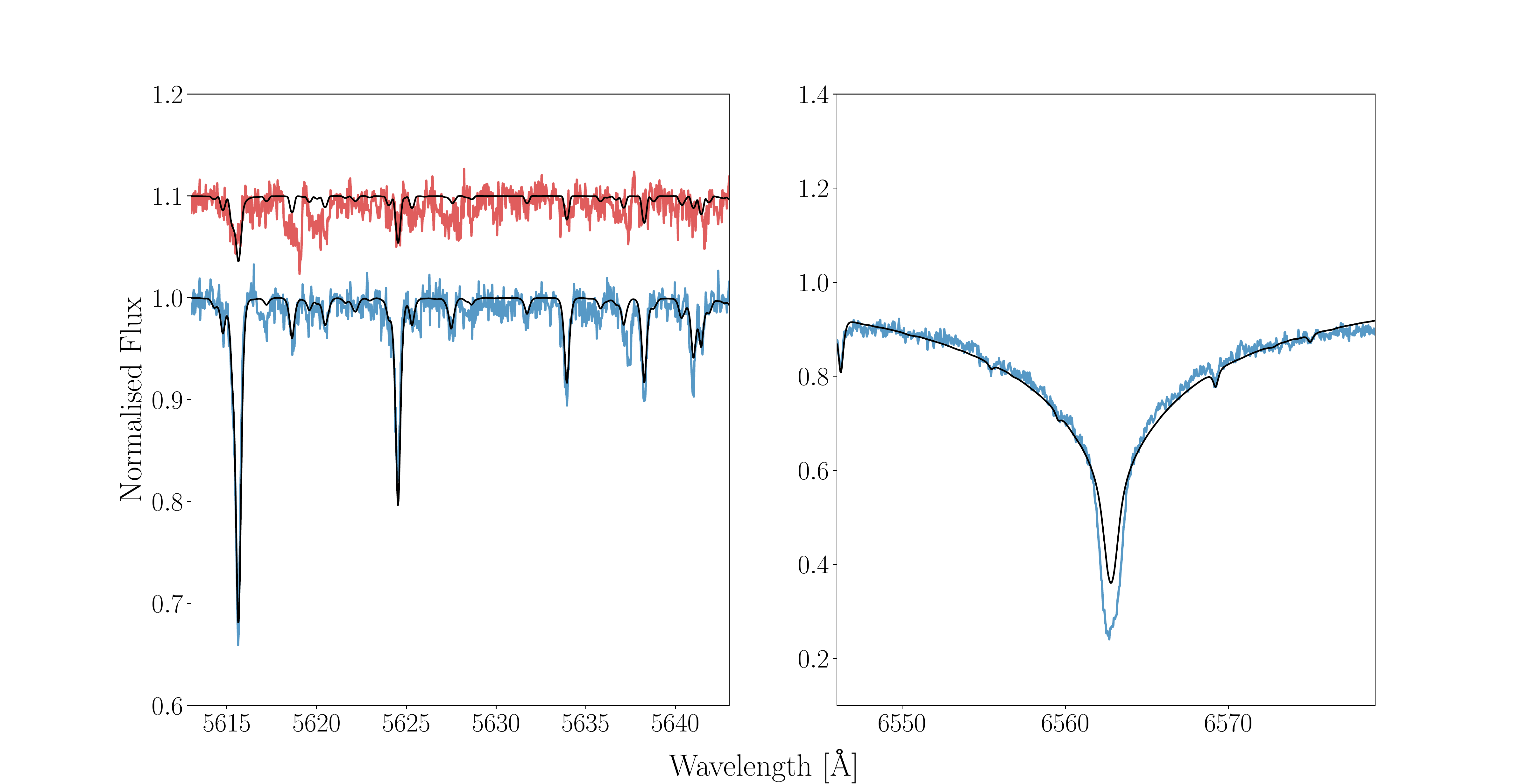}
\caption[Best-fitting synthetic spectrum to the disentangled component spectra of KIC9850387.]{Best-fitting synthetic spectrum (in black), generated by \textsc{gssp}, to the disentangled primary component HIRES spectrum (in blue). The disentangled secondary component HIRES spectrum (in red) is also plotted, with a synthetic spectrum generated with input parameters derived from eclipse modelling (listed in Table \ref{tab: parameters}). The left panel is a metal-line region and the H$_{\alpha}$ region in displayed in the right panel.}
\label{fig: GSSP_fit}
\end{figure*}

Table \ref{tab: parameters} lists the maximum-likelihood estimates and 68\% confidence intervals of $T_{\text{eff,p}}$, $\mathrm{[M/H]}$, $v_{\text{micro,p}}$ and $v_{\text{rot,p}}$ sin $i$, along with the parameters derived from eclipse modelling (see Section \ref{subsec: eclipsemodel}). The $T_{\text{eff,p}}=7335_{-85}^{+85}$ K that we obtained is systematically higher than that used by \cite{Zhang2020} ($T_{\text{eff,p}}=6947\pm152$ K; \citealt{Frasca2016}) in their analysis, and we posit that this is due to a combination of the following factors: 1) the lower resolution of the LAMOST spectra from which the spectroscopic parameters were derived; 2) the difference in methodology used (\citealt{Frasca2016} use a grid of low-resolution spectra of real stars to determine their parameters); and 3) fitting the observed spectrum as if the system was a single star and not a binary. The best-fitting synthetic spectrum with respect to a metal-line and the H$_{\alpha}$ region of the disentangled primary component HIRES spectrum is displayed in Figure \ref{fig: GSSP_fit}. We have also plotted a synthetic spectrum generated with input parameters derived from eclipse modelling and the disentangled secondary component HIRES spectrum for the metal-line region, showing that the derived $T_{\text{eff,s}}$, log $g_{\text{s}}$ and $L_{\text{s}}$ values are, at the very least, qualitatively consistent with the morphology of the secondary spectrum.

\section{\textit{Kepler} photometry}
\label{sec: photometry}

KIC9850387 is one of the 'module-3 stars' in the \textit{Kepler} catalogue: One of the 21 CCD modules of the \textit{Kepler} space telescope failed less than a year after launch in January 2010, which means that those stars that happened to fall on the part of the field-of-view covered by the module were not observed. \textit{Kepler} performed 90$^{\circ}$ rotations every $\sim$93 d, which meant that different stars were within the field-of-view of that module in any given quarter. The overall result is that 20\% of all stars in the nominal \textit{Kepler} mission have additional\footnote{All \textit{Kepler} light curves have monthly gaps of several days coinciding with the downlinking of science data from the satellite towards the earth, during which there is a disruption of data collection.} yearly 93-day gaps in their light curves after the failure of the module.

For KIC9850387, this means that the dataset is missing three (Q7, Q11 and Q15) out of the 18 quarters (from Q0 to Q17) worth of data compiled during the nominal \textit{Kepler} mission from May 2009 to May 2013. While this result has minimal consequence for the purposes of eclipse modelling, the consequences in the context of pulsational analysis and interpretation are significant, and we discuss that in detail in Section \ref{sec: pulsations}.

Instead of utilising the detrended SAP fluxes that we used in our preliminary analysis, we chose to extract the light curves of KIC9850387 directly from the pixel data files provided by the MAST (Mikulski Archive for Space Telescopes). In addition, we utilise a custom mask as defined in \cite{Papics2013}: This mask includes a larger number of pixels than the standard mask, reducing the effects of systematic instrumental trends in the extracted fluxes. Consequently, the amount of detrending that has to be applied to the extracted fluxes is significantly reduced, thereby reducing the potential impact of said detrending in the low-frequency regime and therefore, on the $g$ modes themselves. The remaining systematic trends in each quarter were then corrected by applying a second-order polynomial to the extracted fluxes (as performed in e.g. \citealt{Tkachenko2013a,Debosscher2013,Schmid2015}). 

\subsection{Eclipse modelling setup}
\label{subsec: eclipsemodel}

The dominant source of variability in the light curve of KIC9850387 is its prominent eclipses, enabling the extraction of the fundamental properties of each component when combined with spectroscopic data. To accurately model these eclipses, we utilised a genetic algorithm written in python (Abdul-Masih et al. 2020 in prep; based on \citealt{Charbonneau1995}) wrapped around the state-of-the-art PHOEBE2.0 code (version 2.1.15, \citealt{Prsa2016,Horvat2018}) in order to generate and fit a binary model to our observations. PHOEBE2.0 includes a whole suite of improved physics including 1) a triangular discretisation of stellar surfaces; 2) a robust treatment of reflection and heat redistribution through the inclusion of Lambertian scattering (see \citealt{Prsa2016} for more details); and 3) an improved treatment of limb darkening by interpolating emergent intensities directly from a grid of \cite{Castelli2004} atmospheric models, rather than the standard practice of assuming a parametric limb-darkening function and interpolating coefficients from tables of coefficients (e.g. from \citealt{Claret2011}).


\begin{table}[ht!]
\setlength{\tabcolsep}{12pt}
\renewcommand{\arraystretch}{1.3}
\caption[Systemic, primary, and secondary parameters of KIC9850387.]{Systemic, primary, and secondary parameters of KIC9850387, derived through a combination of atmospheric (see Section \ref{subsec: atm}) and eclipse modelling (see Section \ref{subsec: eclipsemodel}). The errors quoted are based on the 68\% confidence intervals of the fit parameters. Free parameters in eclipse modelling are indicated by $^{*}$.}
\begin{center}
\begin{tabular}{ ccc }
 \hline
 \hline
  Parameter & \multicolumn{2}{c}{Systemic}\\ 
 \hline
 {$P_{\text{orb}}$ (d)} & \multicolumn{2}{c}{$^{*}$2.7484939$_{-0.0000004}^{+0.0000007}$}\\ 
$\omega_{0}$ ($^{\circ}$) & \multicolumn{2}{c}{$^{*}$270$_{-3}^{+9}$}\\ 
$T_{0}$ (d) & \multicolumn{2}{c}{$^{*}$2454956.4185$_{-0.0002}^{+0.0001}$}\\ 
$e$ & \multicolumn{2}{c}{$^{*}$0.0030$_{-0.0019}^{+0.0005}$}\\ 
$i$ ($^{\circ}$) & \multicolumn{2}{c}{$^{*}$82.21$_{-0.02}^{+0.02}$}\\ 
$v_{\gamma}$ (km~s$^{-1}$) & \multicolumn{2}{c}{$^{*}$5.54$_{-0.09}^{+0.04}$}\\ 
{[M/H] (dex)} & \multicolumn{2}{c}{-0.11$_{-0.06}^{+0.06}$}\\ 
 \hline
  Parameter & \multicolumn{1}{c}{Primary} & \multicolumn{1}{c}{Secondary}\\ 
 \hline
 $M$ (M$_\odot$) & $^{*}$1.66$_{-0.01}^{+0.01}$ & 1.062$_{-0.005}^{+0.003}$\\ 
$R$ (R$_\odot$) & $^{*}$2.154$_{-0.004}^{+0.002}$ & $^{*}$1.081$_{-0.002}^{+0.003}$\\ 
$T_{\text{eff}}$ (K) & 7335$_{-85}^{+85}$ & $^{*}$6160$_{-77}^{+76}$\\ 
log $g$ (dex) & 3.992$_{-0.003}^{+0.003}$ & 4.396$_{-0.003}^{+0.003}$\\ 
$v_{\text{micro}}$ (km~s$^{-1}$) & 2.4$_{-0.3}^{+0.3}$ & --\\
$v_{\text{rot}}$ sin $i$ (km~s$^{-1}$) & 13.4$_{-0.8}^{+0.8}$ & --\\
$f_{\text{rot,surf}}$ (d$^{-1}$) & 0.122$_{-0.008}^{+0.008}$ & --\\ 
$K$ (km~s$^{-1}$) & $^{*}$88.7$_{-1.3}^{+0.4}$ & $^{*}$138.9$_{-1.7}^{+0.5}$\\ 
$I_{\text{refl}}$ & $^{*}$0.5$_{-0.4}^{+0.5}$ & $^{*}$0.3$_{-0.3}^{+0.7}$\\ 
$\beta$ & $^{*}$0.46$_{-0.02}^{+0.05}$ & $^{*}$0.8$_{-0.8}^{+0.1}$\\ 
$L_{\text{r}}$ & $^{*}$0.893$_{-0.008}^{+0.002}$ & 0.107$_{-0.008}^{+0.002}$\\ 
 \hline
\end{tabular}
\tablefoot{\\
\noindent Systemic parameters
\begin{itemize}[topsep=0pt]
\item{$P_{\text{orb}}$: Orbital Period}
\item{$\omega_{0}$: Argument of periastron}
\item{$T_{0}$: Time of superior conjunction}
\item{$e$: Orbital eccentricity}
\item{$i$: Orbital inclination}
\item{$v_{\gamma}$: Systemic velocity}
\item{[M/H]: Metallicity}
\end{itemize}
\noindent Primary and Secondary component parameters
\begin{itemize}[topsep=0pt]
\item{$M$: Mass}
\item{$R$: Equivalent radius (the radius that each star would have if it was a perfect sphere)}
\item{log $g$: Logarithm of the surface gravity}
\item{$v_{\text{micro}}$: Microturbulent velocity}
\item{$v_{\text{rot}}$ sin $i$: Projected rotational velocity}
\item{$f_{\text{rot,surf}}$: Surface rotational frequency}
\item{$K$: Radial velocity semi-amplitude}
\item{$T_{\text{eff}}$: Effective temperature}
\item{$I_{\text{refl}}$: Fraction of incident radiation that is reflected by the star}
\item{$\beta$: Gravity darkening exponent\footnote{The gravity darkening exponent determines the degree to which the temperatures (and therefore fluxes) of each surface element of the PHOEBE2.0 model are affected by the surface gravity at that element, according to the relation ${T_{\text{eff}}\propto g^{\beta/4}}$. See \cite{Prsa2016} for a detailed description of its implementation in PHOEBE2.0.}}
\item{$L_{\text{r}}$: Light contribution of the star with respect to the total flux}
\end{itemize}
}
\end{center}
\label{tab: parameters}
\vspace{36pt}
\end{table}

\begin{figure}[b]
\includegraphics[width=\hsize]{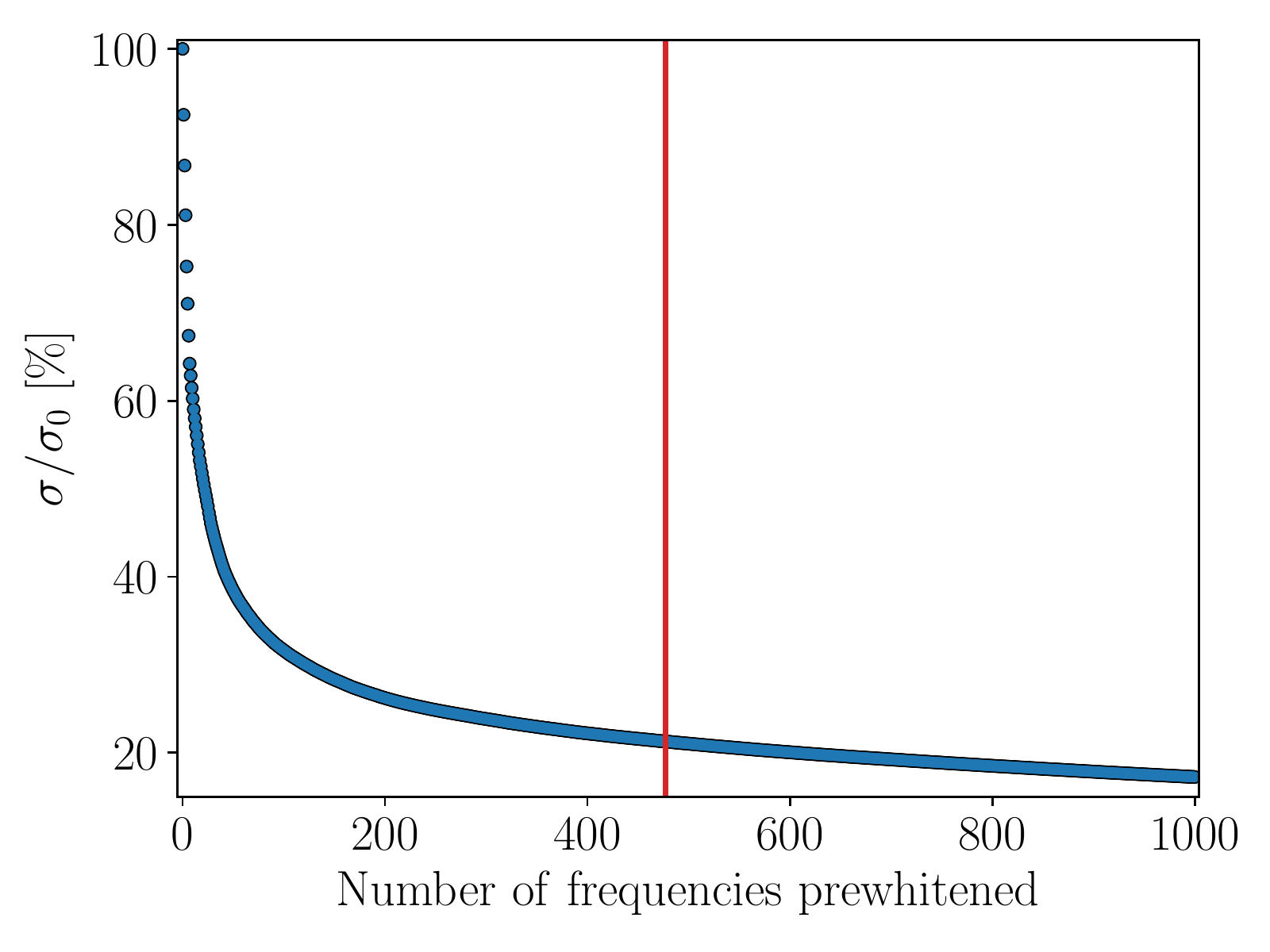}
\caption[Percentage decrease in standard deviation of the light curve with the number of prewhitened frequencies.]{Percentage decrease in standard deviation of the light curve of KIC9850387 with the number of prewhitened frequencies. The vertical red line represents the number of frequencies after which the prewhitening was stopped, based on an iteration-to-iteration relative standard deviation reduction.}
\label{fig: std_reduction}
\end{figure}

Most importantly, PHOEBE2.0 provides model outputs in the form of both photometric fluxes and radial velocities, and hence its inclusion in our framework allows for the simultaneous fitting of the \textit{Kepler} photometric fluxes as well as the radial velocities. Due to the extensive computation time required for each synthetic data point (whether fluxes or velocities), computing synthetic fluxes corresponding to each of the 52~757 observed fluxes ($\sim$420 orbital cycles) of the Kepler light curve would be impractical. Each PHOEBE2.0 eclipse model is therefore constructed by first computing 140 fluxes across a single orbital phase, and then interpolating through these 140 synthetic fluxes to obtain the 52~757 model fluxes corresponding to each observed flux value of the phase-folded light curve. Simultaneously, synthetic radial velocities were computed to match each of the 24 observed radial velocities, and we fit them independent of our previous spectroscopic orbital analysis described in Section \ref{subsec: orbital}. The efficiency of the genetic algorithm was of critical importance as we utilised 17 free parameters in our PHOEBE2.0 models. Through these 17 fit parameters, we are able to derive a total of 21 characteristic systemic, primary and secondary parameters of KIC9850387 (see Table \ref{tab: parameters})

Parameter and error estimation is then performed in the vein of \cite{AbdulMasih2019} by converting the $\chi^{2}$ values of each model output from the genetic algorithm into a probability, using the following methodology \citep{Tramper2011}: 1) Normalise the $\chi^{2}$ values according to the following equation: ${\chi_{\text{norm}}^{2}=(\chi^{2}/\chi_{\text{min}}^{2})*\nu}$, where $\chi_{\text{norm}}^{2}$ is the normalised $\chi^{2}$, $\chi_{\text{min}}^{2}$ is the minimum $\chi^{2}$ value and $v$ is the number of degrees of freedom. This makes it such that normalised reduced $\chi^{2}$, $\chi_{\text{norm, min}}^{2}/\nu=1$. The implicit assumption being made is that the best-fitting model provides a good fit to the data \citep{Tramper2011}, which is not necessarily true in general as this depends on our initial parameter space. This is mitigated to a large extent by our iterative approach (described in Section \ref{subsec: iterative}), enabling the optimisation of the parameter space between iterations; 2) Convert the $\chi_{\text{norm}}^{2}$ values into probabilities using an incomplete gamma function as follows: $P=1-\Gamma(\chi_{\text{norm}}^{2}/2,\nu/2)$; 3) Construct 68\% confidence intervals for each parameter by considering all models with $P>0.32$.

\begin{figure*}[t]
\includegraphics[width=\hsize]{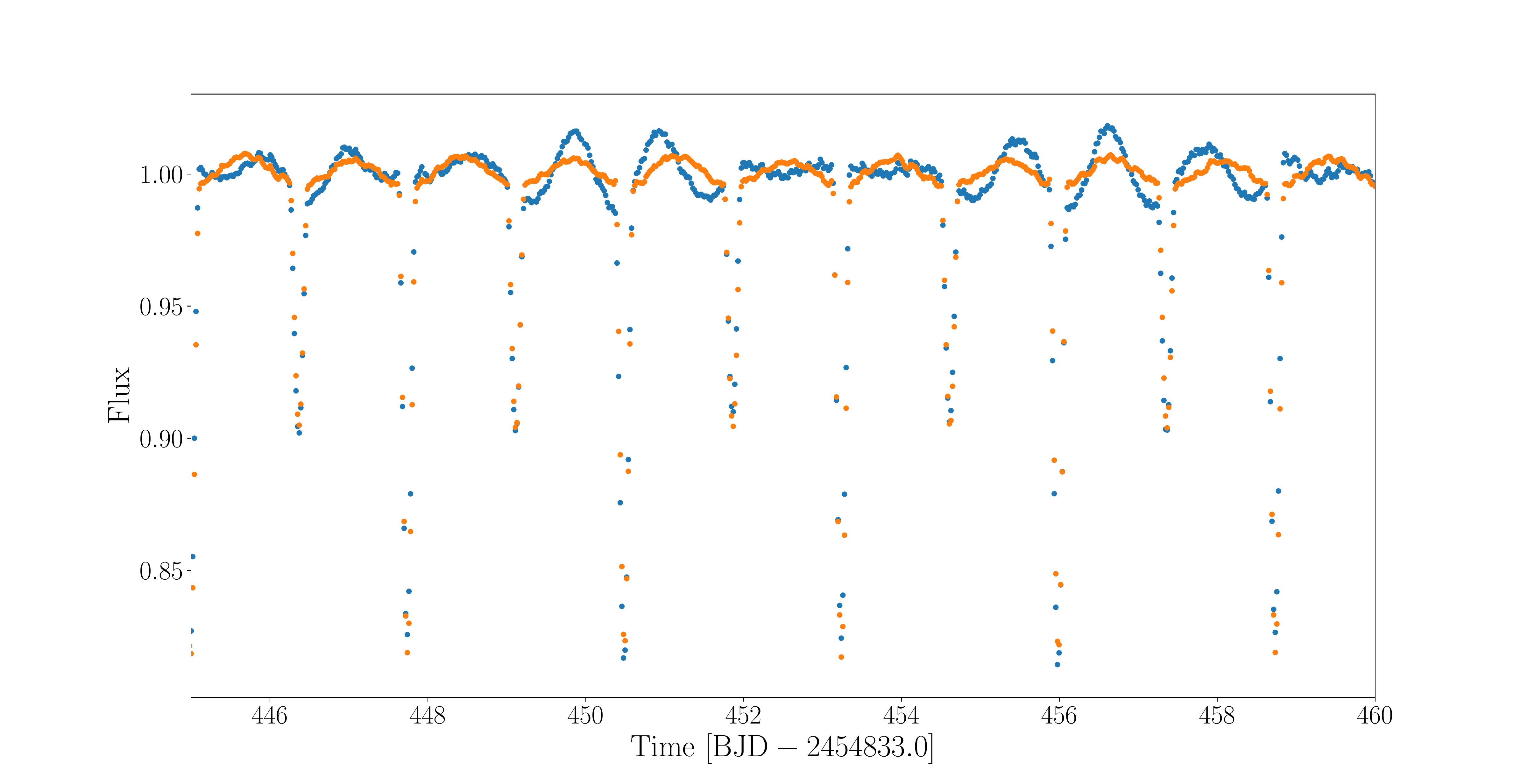}
\caption[15-day section of the light curve of KIC9850387.]{15-day section of the light curve of KIC9850387, showing the original light curve (in blue) and after 478 pulsational frequencies have been removed (in orange).}
\label{fig: LC}
\end{figure*}

\subsection{Obtaining a robust binary model}
\label{subsec: iterative}

The presence of large-amplitude pulsational variability complicates the eclipse modelling process: \cite{Debosscher2013} had noted that both types of variability would have to be disentangled in order to produce a robust binary model. Furthermore, large-amplitude ellipsoidal variation can be observed in the light curve, necessitating the use of a higher order binary-model physics such as gravitational distortion and irradiation to reproduce the observations. Subtracting a model composed of polynomials (e.g. the polyfit models described in \citealt{Prsa2011}) or a binned model of the light curve (e.g. \citealt{Li2020a}) may inadvertently result in the removal of pulsational variation, and as such we chose not to adopt such techniques. 

We therefore adopt the iterative approach in the vein of studies such as \cite{Maceroni2013} and \cite{Debosscher2013}. This methodology is coupled with the iterative methodology for the determination of spectroscopic parameters detailed in Section \ref{subsec: atm}: 1) Generate a pulsational model by iteratively prewhitening (see \citealt{Degroote2009} for a detailed description of the method) the light curve (up to the Nyquist frequency of 24.47 d$^{-1}$) after clipping the eclipses and interpolating through the gaps with cubic splines. We prewhiten the light curve in decreasing order of amplitude of the frequencies in the Lomb-Scargle periodogram \citep{Scargle1982} until the standard deviation in the residual light curve decreases by less than 0.05\% between subsequent prewhitening iterations; 2) Remove the pulsational model from the unaltered light curve, and derive the best-fitting PHOEBE2.0 model for the residual light curve, fixing the $T_{\text{eff,p}}$ at the spectroscopic estimate. Our genetic algorithm setup uses a population of 256 models and is run for 1000 generations, resulting in a total of 256\ 000 model computations in each iteration; 3) Remove the best-fitting PHOEBE2.0 model from the unaltered light curve, and generate a new pulsational model; 4) Perform iterations of eclipse model and pulsational model determination until the difference in the $\chi^{2}$ value of the best-fitting PHOEBE2.0 model between consecutive iterations changes by less than 1\%. Once this point is reached, we retain the eclipse-modelling parameters of the final iteration, and calculate a final pulsational model after removing the best-fitting eclipse model from the unaltered light curve.

\begin{figure}[!b]
\includegraphics[width=\hsize]{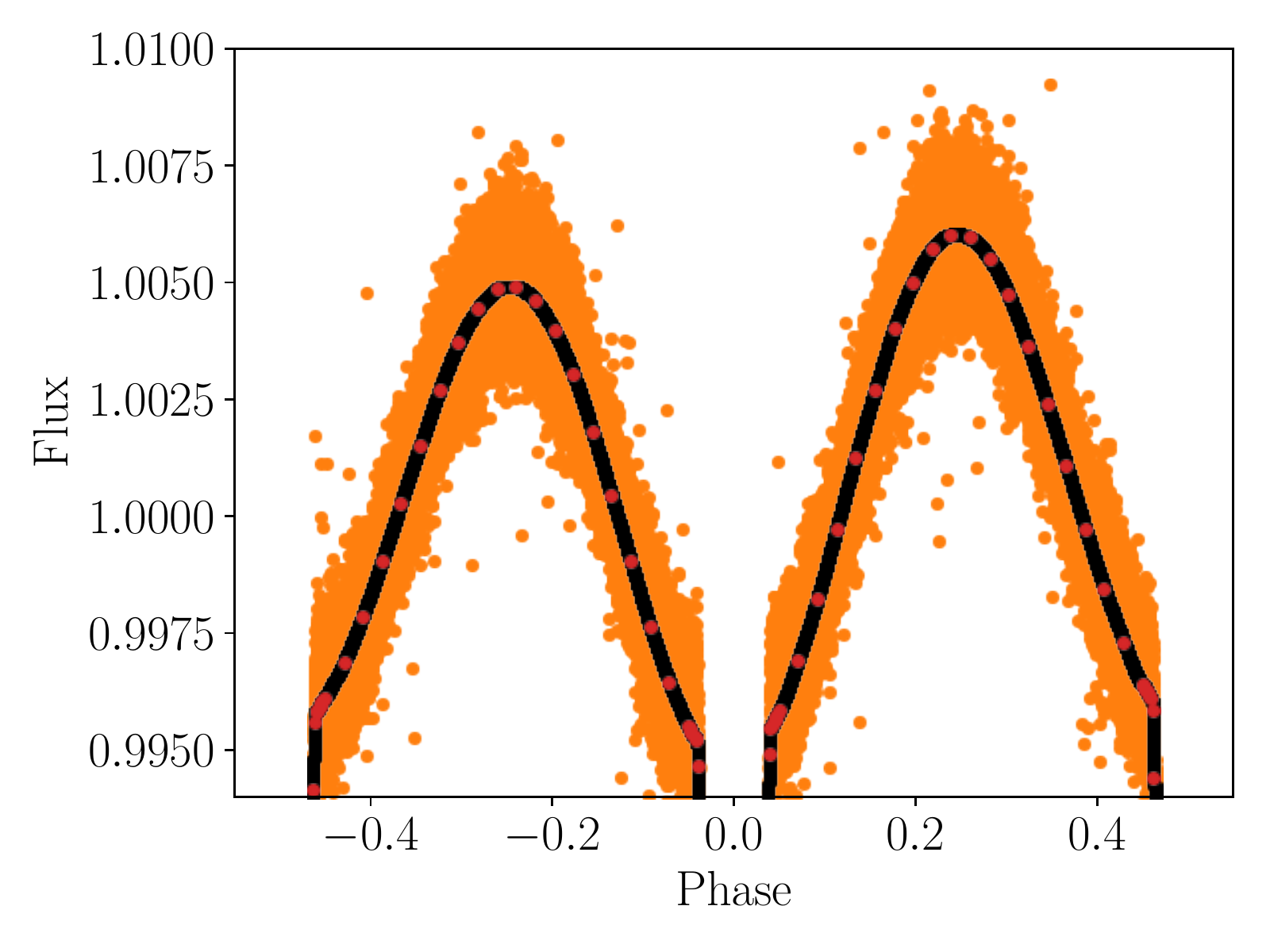}
\caption[Observed asymmetry in the out-of-eclipse light curve.]{Observed asymmetry in the out-of-eclipse light curve with the pulsations removed (in orange), compared to our PHOEBE2.0 models with Doppler boosting included (in black). The red points correspond to the synthetic fluxes that were interpolated through to create the complete eclipse model.}
\label{fig: Eclipse_DB}
\end{figure}

While the oft-quoted procedure in the literature in terms of frequency extraction is to prewhiten the light curve until a frequency is extracted below a S/N cutoff of 4 \citep{Breger1993}, this cutoff is highly dependent on the manner in which the S/N is calculated. One of the ways in which the S/N is calculated is based on a frequency window around each extracted frequency (e.g. \citealt{Tkachenko2013a}) and therefore varies with the density of peaks in that frequency region, as well as with the number of frequencies that have been prewhitened up to that point. Therefore, depending on the star and the adopted S/N calculation methodology, anything from tens to thousands of frequencies would be required before that level is reached. For example, \cite{Debosscher2013} prewhitened over 6000 frequencies, and \cite{Tkachenko2013a} had to reduce their S/N criterion for several of their stars as they found significant residual variation. It was also noted by \cite{Li2019a} that adopting the S/N cutoff of 3 instead of 4 resulted in the extraction of additional frequencies that formed parts of their identified period-spacing patterns. 

Another issue that arises when considering the prewhitening cutoff is that some binary parameters are not well-constrained if there is significant residual variation in the light curve (as noted by \citealt{Debosscher2013}): We discovered that adopting the S/N cutoff of 4, where the noise level was determined from a 1 d$^{-1}$ window around each peak (as we did for our sample characterisation in Section \ref{sec: targets}), resulted the gravity darkening exponent ($\beta$) being poorly constrained during eclipse modelling. We therefore adopted an iteration-to-iteration relative standard deviation reduction (inspired by the approaches in \citealt{Papics2012b}) of 0.05\% as our prewhitening cutoff. We found that this represented a good compromise between removing pulsational variation from the light curve for eclipse modelling while resulting in the extraction of a conservatively high number of frequencies ($\sim$500) in each iteration, making it likely that all frequencies of asteroseismic potential were extracted.

As noted by \cite{Balona2014}, excessive prewhitening of a light curve results in the extraction of spurious frequencies, and therefore the interpretation of the extracted frequencies must be performed with caution. Only a handful to several tens of frequencies have significant asteroseismic value, and a fraction of the frequencies extracted from the light curves of heat-driven pulsators are combination frequencies (see \citealt{Papics2012} and \citealt{Kurtz2015b} for detailed discussions of the origins and interpretations of combination frequencies). While these combination frequencies are ideal for the studying of non-linear effects in pulsations, their interpretation is outside of the scope of this paper. Figure \ref{fig: std_reduction} shows the percentage decrease in standard deviation of the light curve with the number of extracted frequencies (up to 1000). A 15-day section of the original light curve and the light curve after removing 478 frequencies in our final iteration is shown in Figure \ref{fig: LC}.


It was discovered after a few iterations that there was significant out-of-eclipse variability in the residual light curve after eclipse-model removal in the positive half-phase of the orbit (i.e. after each primary eclipse and before each secondary eclipse). Although this could be a result of spot modulation, the asynchronous nature of the binary with respect to the primary star (i.e. the surface rotation rate of the primary star ${f_{\text{rot,surf(p)}}=0.122}$ d$^{-1}$ is close to 1/3 of the orbital frequency ${f_{\text{orb}}=0.364}$ d$^{-1}$) makes this scenario unlikely. We therefore concluded that Doppler boosting (see e.g. \citealt{Bloemen2012}) is the likely mechanism behind this phenomenon, and we included in our models for subsequent iterations. The clear asymmetry between each half-phase of the out-of-eclipse light curve (with the pulsations removed) is shown in Figure \ref{fig: Eclipse_DB}, and the fit of our models was substantially improved.

Our iterative process allowed for the simultaneous optimisation of the eclipse model and dynamical parameters, as well as of the pulsational frequencies that are analysed in Section \ref{sec: pulsations}. The maximum-likelihood estimates and 68\% confidence intervals of our parameters from eclipse modelling are listed in Table \ref{tab: parameters}, along with the spectroscopic parameters (see Section \ref{subsec: atm}). There are significant differences between our parameters and those derived by \cite{Zhang2020}, which is unsurprising considering that they only used the radial velocities of the primary star (with fewer measurements), a different effective temperature of the primary star, and fixed $P_{\text{orb}}$ at the KEBC value and $e=0$. However, their orbital inclination ($i=82.25\pm0.03$), and their log~$g$ values for the primary ($3.98\pm0.03$) and secondary ($4.34\pm0.03$) star are in good agreement with those that we derived.  

It can be seen that $I_{\text{refl}}$ is completely degenerate for both components, implying that the effect of reflection is rather weak. However, $\beta$ was found to be well-constrained for the primary star, and its value of 0.460 is in between the classical theoretical values derived for fully radiative envelopes (1.0; \citealt{VonZeipel1924}), and fully convective envelopes (0.32; \citealt{Lucy1967}). However, according to \cite{ER2012}, $\beta$ is theoretically expected to vary with the amount of ellipsoidal variation of the star, decreasing from 1.0 to 0.8 with increasing degrees of ellipsoidal variation regardless of the type of stellar envelope. They posit that low values of $\beta$ that have been reported (e.g. in \citealt{Djurasevic2003,Djurasevic2006}) are the result of physical effects such as irradiation and asychronous rotation weakening the correlation between $T_{\text{eff}}$ and $g$. Based on these arguments, we can only conclude that there is a weak correlation between $T_{\text{eff}}$ and $g$ and posit that this may be due to the asynchronous rotation of the star and residual pulsational variation in the modelled light curve.

\begin{figure*}[!t]
\includegraphics[width=\hsize]{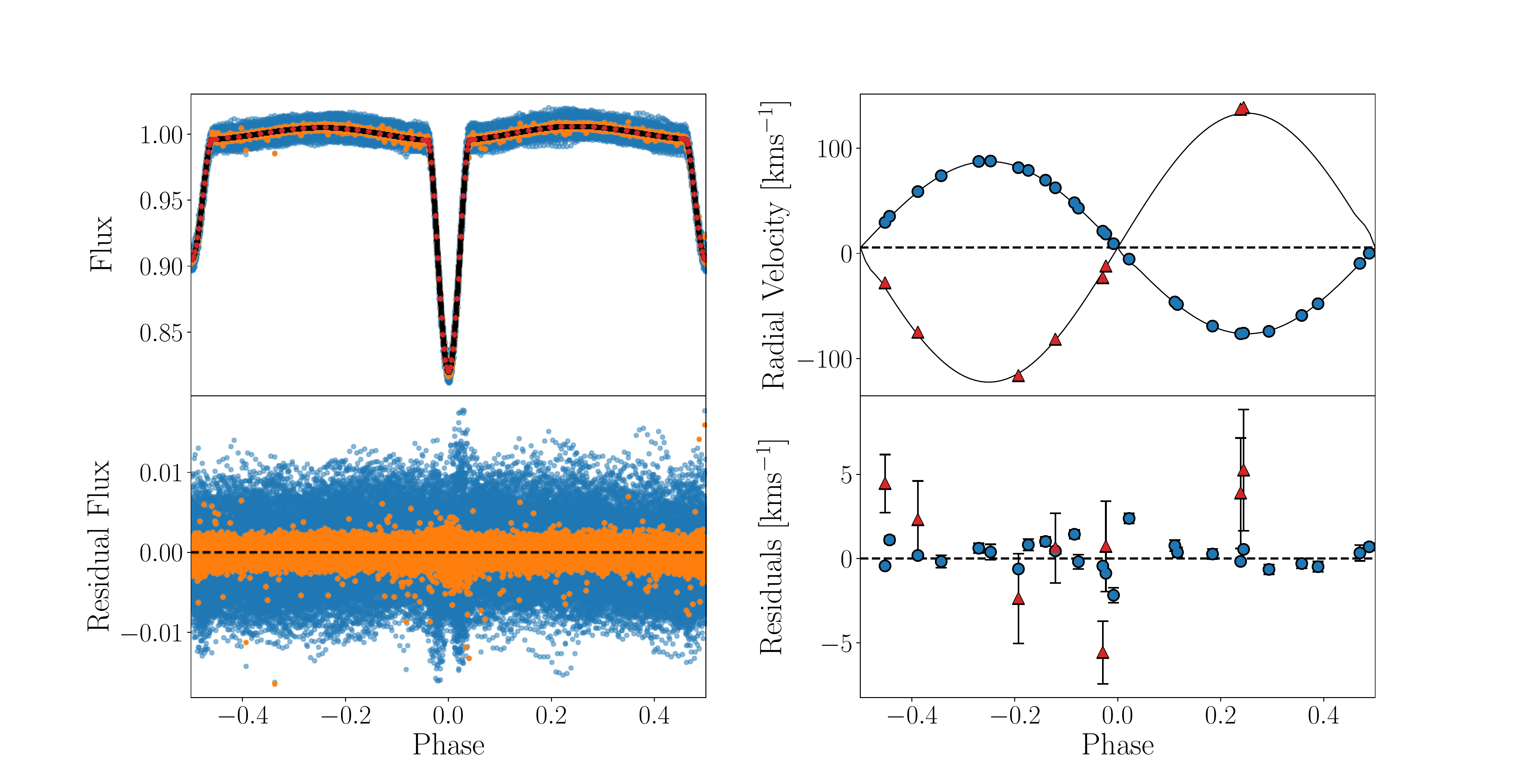}
\caption[Best-fitting PHOEBE2.0 model to our observational data.]{Best-fitting PHOEBE2.0 model to our observational data. The top left panel displays the best-fitting eclipse model (in black) to the unaltered light curve (in blue) and the light curve with the pulsations removed (in orange). The red points correspond to the synthetic fluxes that were interpolated through to create the complete eclipse model. The top right panel displays the best-fitting Keplerian orbital fit to the primary (in blue) and secondary (in red) radial velocities. The residuals of the fit of the eclipse model and the radial velocities are displayed in the bottom left and bottom right panels respectively, with the same colour-coding as the top panels.}
\label{fig: EM_RV_model}

\vspace{18pt}

\includegraphics[width=\hsize]{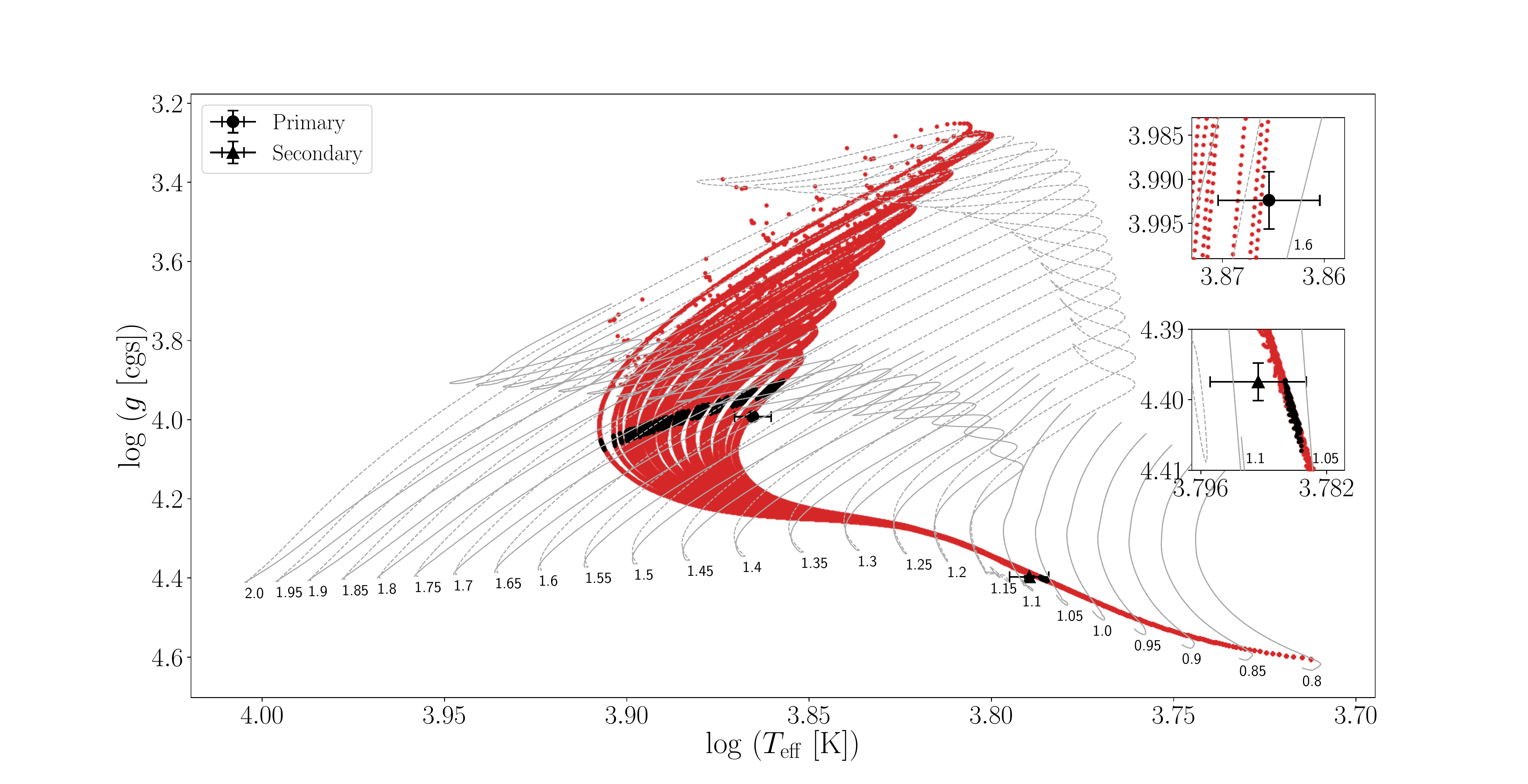}
\caption[Positions of the components of KIC9850387 on the $\text{log }T_{\text{eff}}-\text{log }g$ diagram.]{Positions of the components of KIC9850387 on the $\text{log }T_{\text{eff}}-\text{log }g$ diagram. The red regions corresponds to the best-fitting isochrone cloud \citep{Johnston2019b} to the dynamical parameters of both components. The black regions on the isochrone-cloud represent the $\text{log }T_{\text{eff}}-\text{log }g$ values corresponding to the dynamical masses of the individual components. The grey curves are \textsc{mesa} evolutionary tracks with different amounts of core-boundary mixing (solid and dashed), with their corresponding masses (in units of $M_{\odot}$) indicated at the base of each track. The inset plots are magnified regions around the position of the primary (top) and secondary (bottom) component.}
\label{fig: isocloud_fit}
\end{figure*}

Our best-fitting eclipse model and Keplerian orbital fit from PHOEBE2.0 is displayed in Figure \ref{fig: EM_RV_model}, and we find also good agreement between the $K_{1}$ and $K_{2}$ values derived from our preliminary spectroscopic orbital fit (see Section \ref{subsec: orbital}) and those derived through this analysis. However, there is increased variation in the in-eclipse phases of the residuals (particularly around the primary eclipse) compared to the out-of-eclipse phases, and this phenomenon manifests in the Lomb-Scargle periodogram, as described in Section \ref{sec: pulsations}.

Figure \ref{fig: isocloud_fit} shows the positions of the components of KIC9850387 on the $\text{log }T_{\text{eff}}-\text{log }g$ (or Kiel) diagram, along with the best-fitting isochrone cloud\footnote{An isochrone cloud is a collection of isochrones with different initial input physics (in this case, different core-boundary and envelope mixing values). See \cite{Johnston2019b} for a detailed description and applications of the isochrone-cloud methodology.} at an age of 1.27 Gyr \citep{Johnston2019b}, generated from single-star evolutionary tracks output from the stellar evolutionary code \textsc{mesa} (revision 10348; \citealt{Paxton2011, Paxton2018}) at an initial metallicity ($Z_{\text{ini}}$)  of 0.010, to the dynamical parameters. One can clearly see that the dynamical parameters of the secondary component are in good agreement with those predicted by evolutionary theory (the black data points) but the primary is not. It was noted by \cite{Tkachenko2020} that the enforcement of binary co-evolution resulted in increased mass discrepancy for several of their stars, and while relaxing the equivalent-age criterion reduces the discrepancy in the primary mass, it does not eliminate it entirely (these parameters only agree within 2-$\sigma$ of the dynamical mass). However, if we also consider evolutionary models calculated at initial metallicities that fall within the spectroscopic errors (i.e. $0.008 \lesssim Z_{\text{ini}} \lesssim 0.012$), this discrepancy disappears. Overall, our results disagree with the claim of \cite{Zhang2020} that the components are pre-main-sequence stars and strongly support a main-sequence binary evolutionary stage. A full description of our isochrone-cloud fitting methodology, asteroseismic modelling and comparisons of dynamical, evolutionary and asteroseismic parameters will be presented in our companion paper, Sekaran et al. (in prep.). 


\begin{figure}[t]
\includegraphics[width=\hsize]{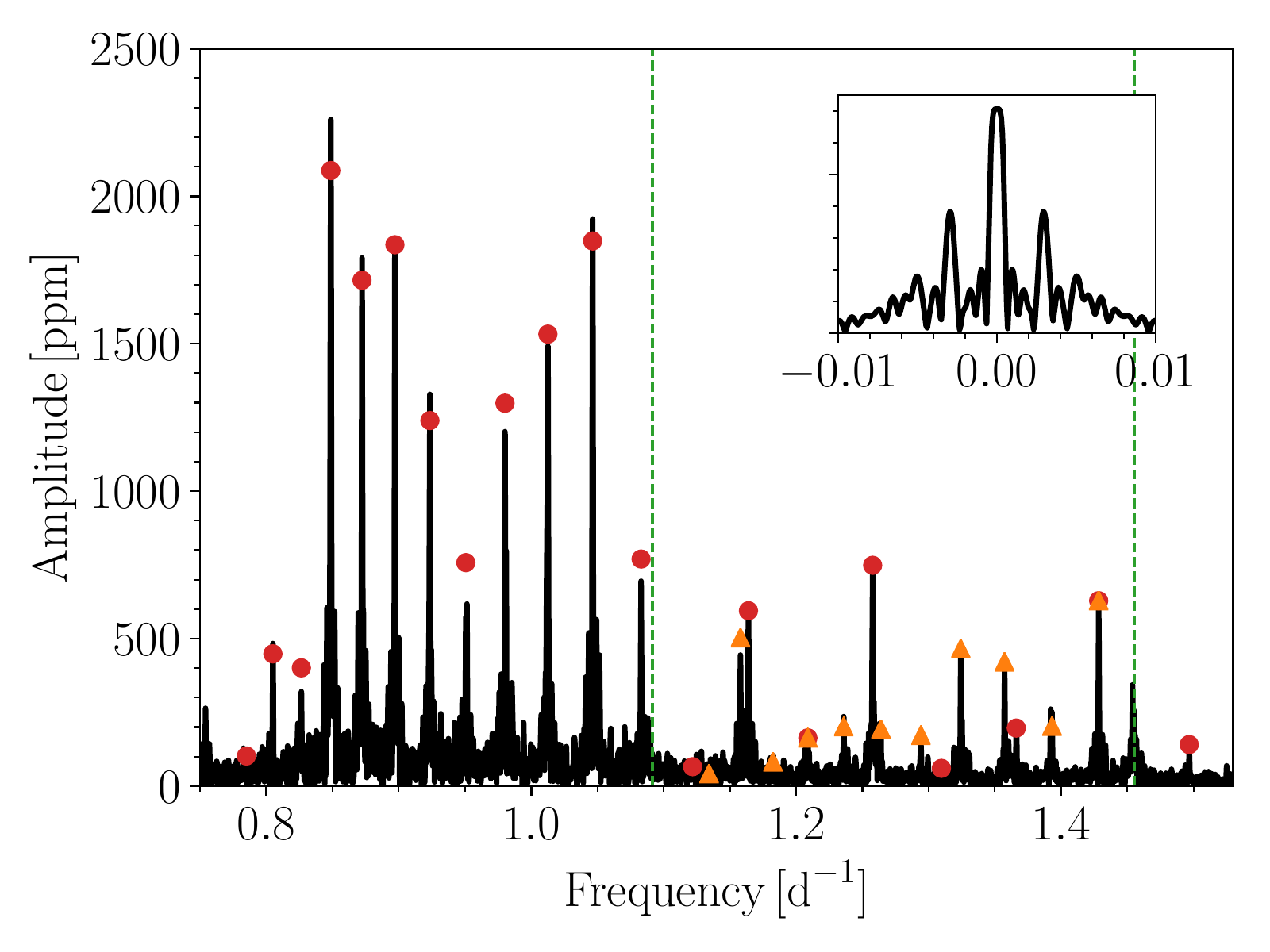}
\caption[Low-frequency region of the Lomb-Scargle periodogram of KIC9850387.]{Low-frequency region of the Lomb-Scargle periodogram of KIC9850387, with the peaks that form the $\ell=1$ and $\ell=2$ period-spacing patterns indicated by red circles and orange triangles, respectively. The vertical green dashed lines represent the orbital harmonics, and the inset plot shows the spectral window.}
\label{fig: PS_SW_FT}

\includegraphics[width=\hsize]{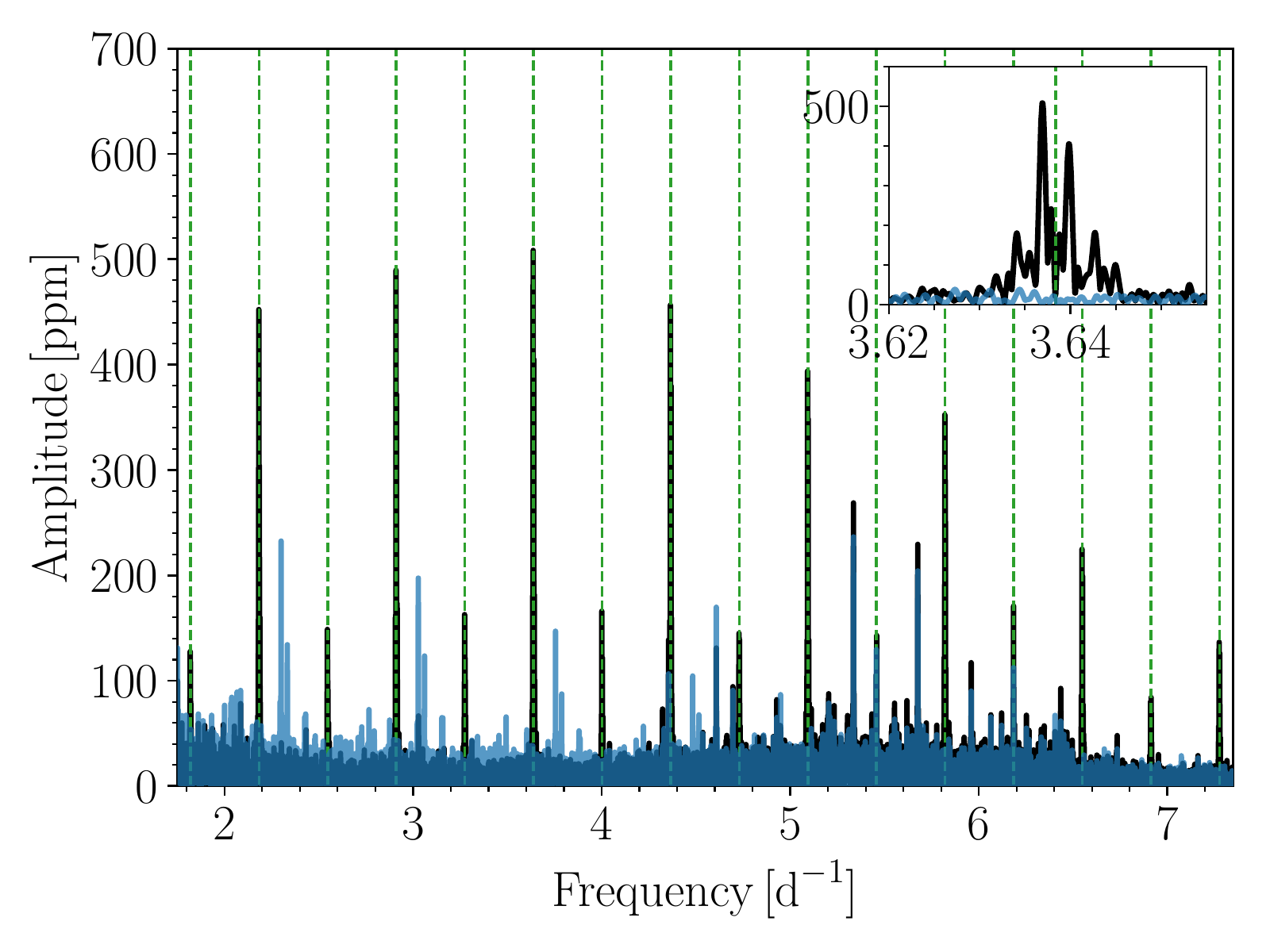}
\caption[Residual peaks around orbital harmonics in the Lomb-Scargle periodogram of KIC9850387.]{Residual peaks around orbital harmonics (vertical green dashed lines) in the Lomb-Scargle periodogram of KIC9850387. The periodogram of the original light curve is shown in black, and the periodogram of the light curve after the eclipse regions were clipped and interpolated through with cubic splines is shown in blue. The inset plot shows a magnified region of the frequency multiplet with the highest amplitude.}
\label{fig: ER_FT}
\end{figure}

\section{Asteroseismic analysis}
\label{sec: pulsations}

The first step in the asteroseismic analysis is to remove frequencies that are within a certain resolution criterion of another higher-amplitude frequency. As mentioned in Section \ref{sec: photometry}, we extracted more than the standard (as per the \citealt{Breger1993} criterion) amount of frequencies to optimise the binary model, and as such we would need to identify the independent pulsational frequencies. The standard practice in these instances is to remove any frequency that is within a multiple (1 or 1.5 times) of the Rayleigh resolution ($R=1/\Delta\tau$) of another \citep{Degroote2010}, where $\Delta\tau$ is the length of the dataset. This results in a frequency resolution of $R=0.00068$ d$^{-1}$. Due to the additional gaps in the light curve mentioned in Section \ref{sec: photometry}, each frequency peak in the Lomb-Scargle periodogram is split into a multiplet with each peak separated by the \textit{Kepler} orbital frequency of $1/372.5=0.00268$ d$^{-1}$ (see \citealt{Murphy2014, Bowman2016a} for more details). This phenomenon is displayed in Figure \ref{fig: PS_SW_FT}, which shows the low-frequency region of the periodogram where clear series of peaks that could form period-spacing patterns are visible. The inset plot is the spectral window for the periodogram, clearly showing this multiplet phenomenon.

The spectral window shows one prominent peak to either side of the main pulsation frequency separated by the \textit{Kepler} orbital frequency, as well as less-prominent peaks at twice the \textit{Kepler} orbital frequency. We therefore adopted a more conservative resolution criterion of twice the \textit{Kepler} orbital frequency (i.e. $R=0.00536$ d$^{-1}$) such that both the first- and second-order side-peaks are considered when removing frequencies\footnote{It was noted by \cite{Murphy2014} that prewhitening the central frequency of a multiplet does not remove the entire multiplet, and as such justifies our approach.}. In addition, we also removed frequencies below 0.01 d$^{-1}$ as they were likely a consequence of residual ellipsoidal variation in the out-of-eclipse light curve (a phenomenon also noted in \citealt{Maceroni2013}).

In addition, frequency multiplets were observed around each orbital harmonic (see Figure \ref{fig: ER_FT}). It can be seen from the inset plot that these are not exactly at each orbital harmonic but around each orbital harmonic, and the peaks disappear if the light curve was clipped and interpolated between the eclipse regions. We therefore conclude that these peaks are a result of residual variation from incomplete eclipse removal (as mentioned in Section \ref{subsec: eclipsemodel}) and not tidally induced or perturbed pulsational peaks as displayed by pulsating binaries such as U~Gru \citep{Bowman2019} and V453 Cyg \citep{Southworth2020}. As such, we also removed any peaks that were within the adopted resolution criterion ($R=0.00536$ d$^{-1}$) of any orbital harmonic. A total of 193 frequencies remained for further analysis after frequency removal.

\begin{figure}[b]
\includegraphics[width=\hsize]{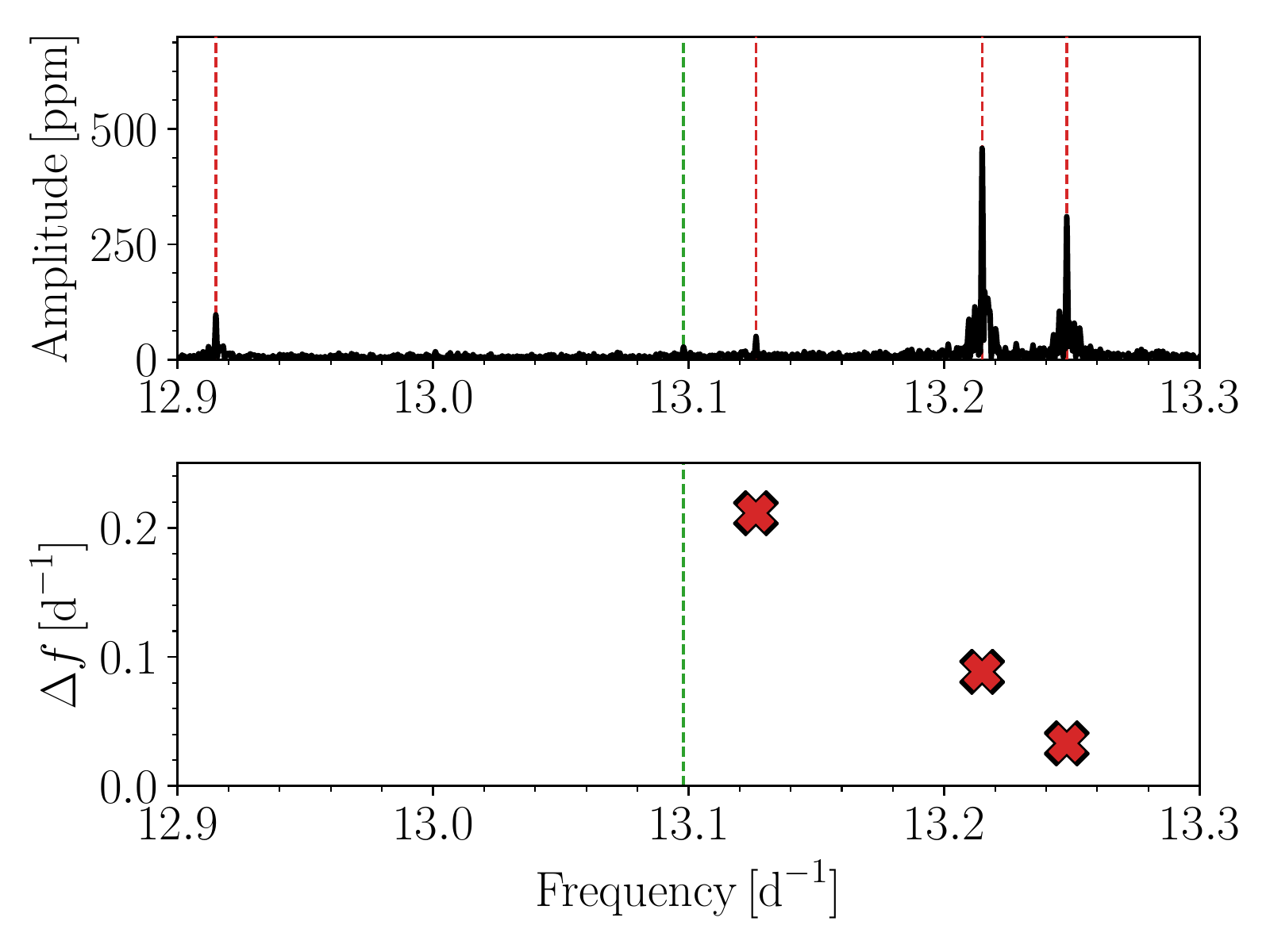}
\caption[Independent $p$-mode frequencies in the Lomb-Scargle periodogram of KIC9850387.]{Independent $p$-mode frequencies (vertical red dashed lines) in the Lomb-Scargle periodogram (in black) of KIC9850387 (top panel). The frequency difference ($\Delta f$) between consecutive $p$-mode frequencies are represented by red 'x' markers (bottom panel). The orbital harmonics in both panels are indicated by vertical green dashed lines.}
\label{fig: pmodes}
\end{figure}

\begin{figure*}[t]
\includegraphics[width=\hsize]{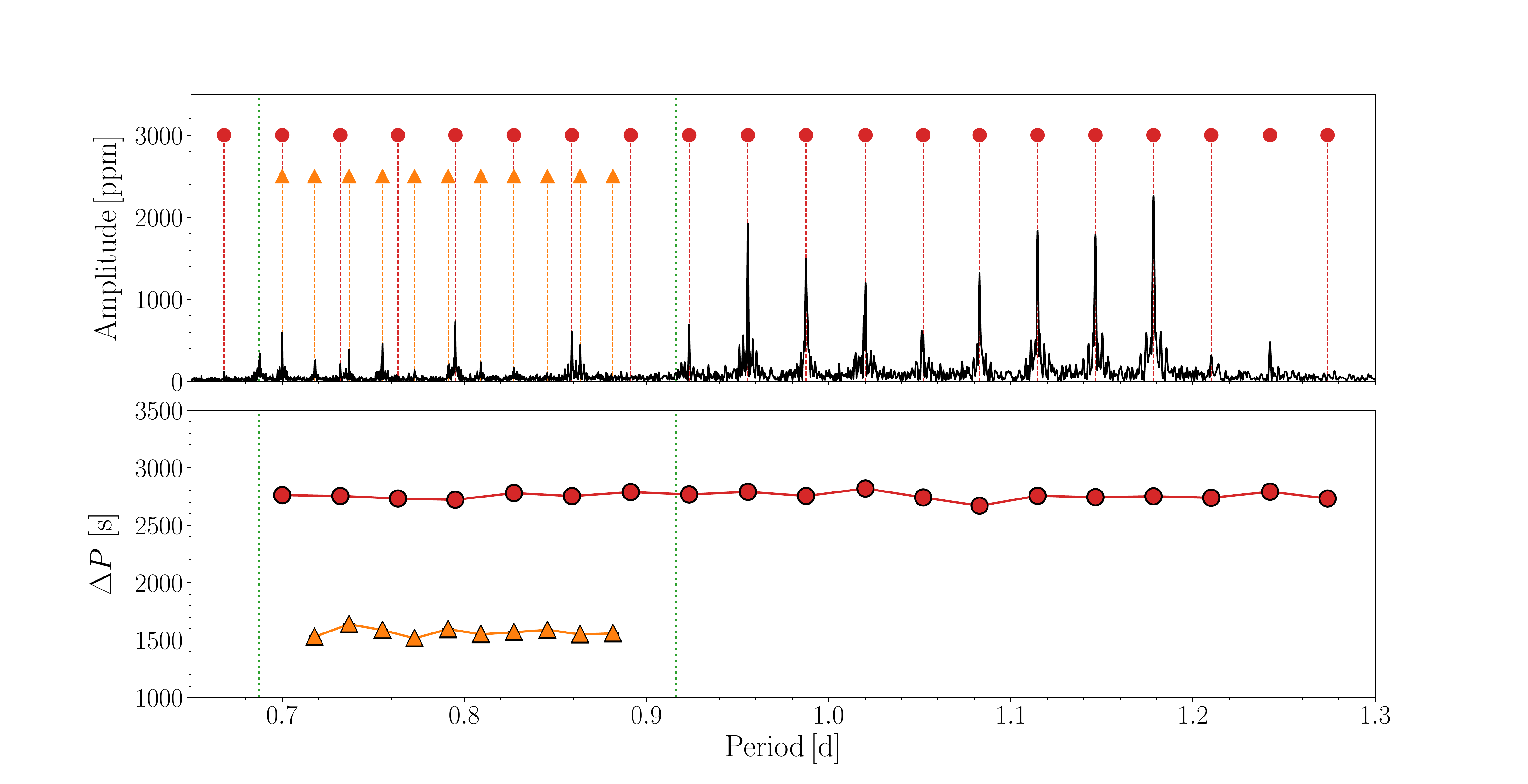}
\caption[Period-spacing patterns of KIC9850387.]{Period-spacing patterns of KIC9850387. The bottom panel shows the $\ell=1$ (red circles) and $\ell=2$ (orange triangles) period-spacing patterns, with the period-spacing on the vertical axis and the period on the horizontal axis. The top panel shows the corresponding periods that were selected to form the pattern, with the Lomb-Scargle periodogram (in black). The orbital harmonics are indicated by vertical green dashed lines in both panels.}
\label{fig: Period_Spacing}
\end{figure*}

The second step in the asteroseismic analysis of any heat-driven pulsator is the identification of combination frequencies. As mentioned in Section \ref{subsec: iterative}, many of the frequencies extracted are potentially mathematical combinations of other frequencies. As discussed in detail in \cite{Papics2012}, some of these frequencies are a result of non-linear interaction between two or more 'parent' frequencies and have a physical interpretation (see e.g. \citealt{Bowman2016b}), while others are simply mathematical artefacts caused by interpreting pulsational frequencies in terms of harmonic functions. Nevertheless, these two types of combinations can be distinguished from each other by considering the phase behaviour of the combination frequencies \citep{Degroote2009} and therefore have to be identified.

We performed our combination frequency search by adopting the methodology of \cite{Papics2012}: We allow for combinations up to the combination order $\mathcal{O}=2$, and consider combinations with frequencies up to the second harmonic. This means that we allow for combinations up to, for example, $2f_{1}\pm2f_{2}$, where $f_{1}$ and $f_{2}$ are two extracted frequencies. We also include the orbital frequency ($f_{\text{orb}}$) in the list of frequencies that we derive combinations of. We consider a frequency to be combination of two other frequencies if 1) it is of lower amplitude than both of the 'parent' frequencies, and 2) if it is within our adopted frequency resolution criterion (i.e. within 0.00536 d$^{-1}$) of the mathematical combination. All of the frequencies ($f$) that were left after removing those that failed our resolution criteria (the first step), and their corresponding errors ($\sigma_{\text{f}}$) are listed in Table \ref{tab: freqlist}, along with their corresponding amplitudes ($A$) and amplitude errors ($\sigma_{\text{A}}$), phases $\phi$ and phase errors ($\sigma_{\phi}$), S/N values, and the combination frequency associated with each (if any). The S/N values that we quote here are computed from the Lomb-Scargle periodogram of the residual light curve after frequency extraction, where the noise level is determined from the mean amplitude of the periodogram in a 1 d$^{-1}$ window centred on each extracted frequency. For the asteroseismic analysis, we retain all independent frequencies with $\text{S/N} > 4$ along with $\text{S/N} > 3$ frequencies that comprise part of a period-spacing pattern. 

Both $p$-mode and $g$-mode frequencies were extracted from the light curve of KIC9850387. Due to the low mass of the secondary ($M_{\text{s}}=1.0476$ M$_{\odot}$), it is highly unlikely that any of the frequencies extracted are due to the pulsations of the secondary star. Stars in the mass-vicinity of 1.0 M$_{\odot}$ tend to pulsate stochastically at very high frequencies well above 3000 $\mu$Hz or 250 d$^{-1}$ \citep{Garcia2019}, which is also well above the Nyquist frequency of 24.47 d$^{-1}$. We therefore conclude that all of the frequencies extracted are a result of pulsations originating in the more-massive primary star.

\subsection{$p$ modes}
\label{subsec: pmodes}

The $p$-mode regime of the Lomb-Scargle periodogram of KIC9850387 (see Figure \ref{fig: pmodes}) is rather sparse, containing just a few independent $p$ modes. No frequency splittings or other characteristic spacing was observed from these frequencies, although the frequency difference ($\Delta f$) between adjacent $p$-mode frequencies decreases as the frequency increases. The surface rotational frequency of $f_{\text{rot,surf(p)}}=0.122$ d$^{-1}$ is very similar to that found by \cite{Schmid2015} for their slowly-rotating F-type pulsating binary KIC10080943, but unfortunately there is no rotational signature in the $p$ modes for this star.

\subsection{$g$-mode period-spacing patterns}
\label{subsec: gmodes}

The Lomb-Scargle periodogram of KIC9850387 features numerous $g$-mode peaks, and we were able to construct two different period-spacing patterns of $\ell=1$ and $\ell=2$ modes. Due to the fact that KIC9850387 is a slow rotator, the mean period-spacing values for these modes should be approximately equal to the asymptotic period-spacing values. Based on this assumption, we obtained $\Pi_{1}\approx2754\pm16$ s and $\Pi_{2}\approx1568\pm12$ s and confirmed that these were indeed an $\ell=1$ and an $\ell=2$ pattern, from the distributions of $\Pi_{\ell}$ values published by \cite{VanReeth2016}. Using Eqs.  (\ref{eq: Piell}) and (\ref{eq: Pinaught}), we also calculated $\Pi_{0}$ for each mode pattern, obtaining $\Pi_{0, \ell=1}=3842\pm29$ s and $\Pi_{0, \ell=2}=3896\pm22$ s, which are within 2$\sigma$ of each other and indicating that both patterns originate from the same star. These values are also consistent with the $\Pi_{0}=3898\pm1$ obtained by \cite{Zhang2020} for their $\ell=1$ pattern, and the $\Pi_{0}=3894\pm7$ s obtained by \cite{Li2020a}.

These patterns comprise at least ten radial orders for each mode, a remarkable discovery in the context of eclipsing binary pulsators, allowing for stringent constraints of stellar structure during asteroseismic modelling (as detailed in \citealt{Schmid2016}). The patterns that we have obtained are longer than those reported by \cite{Li2020a}. Figure \ref{fig: Period_Spacing} shows the period-spacing patterns that were constructed from the $g$-mode frequencies of KIC9850387.

\section{Discussion and conclusions}
\label{sec: conclusions}

In this paper, we detailed the identification and characterisation of a sample of detached eclipsing binaries with excellent $g$-mode asteroseismic potential by performing pulsational screening of all eclipsing binaries in the KEBC between 6000 K and 10 000 K that were considered to be predominantly detached according to \cite{Matijevic2012}. We identified a total of 93 eclipsing binary systems with $g$-mode pulsating components, of which 11 systems contained hybrid $p$- and $g$-mode pulsators. We found clear period-spacing patterns in a total of seven stars, of which two featured continuous patterns longer than six radial orders. We also characterised the pulsations of these 93 eclipsing binary systems with $g$-mode pulsating components by calculating the frequency of highest amplitude ($f_{\text{max}}$) in the $g$-mode regime, and the number of independent frequencies ($N_{\text{ind}}$) in said $g$-mode regime, and compared these parameters with the binary/orbital parameters log $P_{\text{orb}}$, Morph, $T_{\text{eff}}$, $e$, and $\omega$ using the Spearman's rank correlation ($\rho$). The low $|\rho|$ and moderate-to-high $p$-values indicate that the $g$-mode pulsational parameters and binary and atmospheric parameters are weakly correlated at best, as expected for detached main-sequence binaries.

In addition, we presented the observational spectroscopic, photometric and asteroseismic analysis of the pulsating eclipsing binary KIC9850387. First classified as a $\gamma$ Doradus pulsator by \cite{Gaulme2019}, this star was identified during our sample selection and characterisation process as the most promising candidate in our sample for future evolutionary and asteroseismic modelling due to the discovery of multimodal period-spacing patterns in its frequency spectra. We then proceeded with spectroscopic follow-up, compiling a total of 18 HERMES \citep{Raskin2011} and eight HIRES \citep{Vogt1994} spectra. Radial velocities were extracted from these spectra and used to determine the spectroscopic orbital elements, and these elements were then used to perform spectral disentangling. We were able to determine the atmospheric parameters and chemical abundances for the primary star by fitting synthetic spectra. However, due to the low S/N of the disentangled secondary component spectrum, we were only able to obtain qualitative agreement between the disentangled component spectrum and the parameters extracted from the subsequent eclipse modelling process.

We employed an iterative methodology in the vein of studies such as \cite{Maceroni2013} and \cite{Debosscher2013} to simultaneously optimise the pulsational and eclipse models, enabling the extraction of a whole host of parameters including the component masses, radii and the effective temperature of the secondary. We obtained $M_{\text{p}}=1.66_{-0.01}^{+0.01}$ $M_{\odot}$ and $M_{\text{s}}=1.062_{-0.005}^{+0.003}$ $M_{\odot}$, and $R_{\text{p}}=2.154_{-0.004}^{+0.002}$ $R_{\odot}$ and $R_{\text{s}}=1.081_{-0.002}^{+0.003}$ $R_{\odot}$, implying precisions well below the 1\% level. We also obtained $T_{\text{eff,p}}=7335_{-85}^{+85}$ K and $T_{\text{eff,s}}=6160_{-77}^{+76}$ K by iterating between atmospheric and eclipse modelling. We also noted that there is no mass discrepancy for either component, and that the models with the greatest agreement with the observed $T_{\text{eff}}$ and log $g$ tended to have low amounts of core-boundary mixing. As detailed in Sections \ref{sec: spectroscopy} and \ref{sec: photometry}, our results are in general disagreement with those of \cite{Zhang2020}. We found that the system is a SB2 comprising two main-sequence components. The latter results contradict their claims that the system is a SB1 comprising two pre-main-sequence components. We posited that this disagreement is a result of the different quantity and quality of spectra used: We used numerous high-resolution HERMES and HIRES spectra, while \citep{Zhang2020} used only a few lower-resolution LAMOST spectra. Therefore, they were unable to properly characterise the secondary component and subsequently performed eclipse modelling based on insufficient spectroscopic information.

After performing a combination frequency search, we analysed the independent $p$ modes and $g$ modes of the star. The $p$-mode frequency spectrum was sparse with only four independent $p$ modes observed and no frequency splittings or characteristic spacing. Analysis of the rich frequency spectrum of $g$ modes revealed $\ell=1$ and $\ell=2$ period-spacing patterns that were longer than ten radial orders each. \cite{Li2020a} had reported a core rotation rate of 0.0053 d$^{-1}$ from their fitting of the slopes of the $\ell=1$ and $\ell=2$ period-spacing patterns. This is below our adopted resolution criterion of 0.00536~d$^{-1}$ and would render any frequency splitting in the $g$-mode regime indistinguishable from effects of the spectral window of the star (see Figure \ref{fig: PS_SW_FT}). In contrast with our spectroscopic and photometric analysis, our asteroseismic analysis results agree with the conclusion of \cite{Zhang2020} that KIC9850387 is a $\gamma$ Doradus-$\delta$ Scuti hybrid pulsator.

The period-spacing series obtained for this star allow for constraints on the interior mixing profile inferred from evolutionary modelling. As such, we coupled this observational analysis with an evolutionary and asteroseismic modelling-based analysis for the purposes of comparing the observationally and theoretically derived parameters of this star. This theoretical analysis and parameter comparison will be presented in the companion paper Sekaran et al. (in prep.).

\begin{acknowledgements}

The authors would like to thank the anonymous referee who helped us improve the presentation of our results. The research leading to these results has received funding from the Fonds Wetenschappelijk Onderzoek - Vlaanderen (FWO) under the grant agreements G0H5416N (ERC Opvangproject) and G0A2917N  (BlackGEM), and from the European Research Council (ERC) under the European Union’s Horizon 2020 research and innovation programme (grant agreement no. 670519: MAMSIE). MA acknowledges support from the FWO-Odysseus program under project G0F8H6N. DH acknowledges support from the Alfred P. Sloan Foundation and the National Aeronautics and Space Administration (80NSSC19K0597). The authors wish to recognize and acknowledge the very significant cultural role and reverence that the summit of Mauna Kea has always had within the indigenous Hawai‘ian community.  We are most fortunate to have the opportunity to conduct observations from this mountain. The authors would also like to thank the Leuven MAMSIE team for useful discussions. 

\end{acknowledgements}

\defcitealias{Alicavus2014}{AS14}
\defcitealias{Borkovits2014}{Bor14}
\defcitealias{Bradley2015}{Bra15}
\defcitealias{Gaulme2019}{GG19}
\defcitealias{Debosscher2011}{Deb11}
\defcitealias{Debosscher2013}{Deb13}
\defcitealias{Hambleton2013}{Ham13}
\defcitealias{Helminiak2019}{Hel19}
\defcitealias{Kjurkchieva2016}{Kju16}
\defcitealias{Kjurkchieva2017a}{KA17}
\defcitealias{Kjurkchieva2018}{KV18}
\defcitealias{Kurtz2015a}{Kur15a}
\defcitealias{Li2019a}{Li19a}
\defcitealias{Lurie2017}{Lur17}
\defcitealias{Matson2016}{Mat16}
\defcitealias{Rowe2010}{Row10}
\defcitealias{Sowicka2017}{Sow17}
\defcitealias{Uytterhoeven2011}{Uyt11}
\defcitealias{Zhang2018}{Zha18}
\defcitealias{Zhang2020}{Zha20}
\defcitealias{Li2020a}{Li20a}

\bibliographystyle{aa}
\bibliography{KIC9850387}

\begin{appendix}
\onecolumn
\section{Sample of eclipsing binary systems with $g$-mode pulsating components}
\label{sec: sample}

\begin{ThreePartTable}
\setlength{\tabcolsep}{10pt}
\LTcapwidth=\linewidth
\renewcommand{\arraystretch}{1.3}
\renewcommand\TPTminimum{\linewidth}
\begin{TableNotes}
  \item[*]{Period-spacing patterns were found during our pulsational screening process.}
  \item[**]{Period-spacing patterns were found by \cite{Li2020a}.}
  \item{\bfseries References:}
  \item{\citetalias{Alicavus2014}: \cite{Alicavus2014}; \citetalias{Borkovits2014}: \cite{Borkovits2014}; \citetalias{Bradley2015}: \cite{Bradley2015}; \citetalias{Debosscher2011}: \cite{Debosscher2011}; \citetalias{Debosscher2013}: \cite{Debosscher2013}; \citetalias{Gaulme2019}: \cite{Gaulme2019}; \citetalias{Hambleton2013}: \cite{Hambleton2013}; \citetalias{Helminiak2019}: \cite{Helminiak2019}; \citetalias{Kjurkchieva2016}: \cite{Kjurkchieva2016}; \citetalias{Kjurkchieva2017a}: \cite{Kjurkchieva2017a}; \citetalias{Kjurkchieva2018}: \cite{Kjurkchieva2018}; \citetalias{Kurtz2015a}: \cite{Kurtz2015a}; \citetalias{Li2019a}: \cite{Li2019a}; \citetalias{Li2020a}: \cite{Li2020a}; \citetalias{Lurie2017}: \cite{Lurie2017}; \citetalias{Matson2016}: \cite{Matson2016}; \citetalias{Rowe2010}: \cite{Rowe2010}; \citetalias{Sowicka2017}: \cite{Sowicka2017}; \citetalias{Uytterhoeven2011}: \cite{Uytterhoeven2011}; \citetalias{Zhang2018}: \cite{Zhang2018}; \citetalias{Zhang2020}: \cite{Zhang2020}}
\end{TableNotes}
{\small
\begin{longtable}{ccccccccc}
\caption{KIC IDs, pulsational ($N_{\text{ind}}$ and $f_{\text{max}}$) parameters and binary and atmospheric (log $P_{\text{orb}}$, Morph, $T_{\text{eff}}$, $e$, and $\omega$) parameters of the 93 detached eclipsing binary systems with $g$-mode pulsating components identified during our study. The second-to-last column of the table indicates whether $p$-mode frequencies were also observed in their frequency spectra, and the last column lists the truncated references to studies in which $g$ modes were discovered or analysed in these systems.}\label{tab: sample}\\
\hline
\hline
KIC & $N_{\text{ind}}$ & $f_{\mathrm{max}}$ (d$^{-1}$) & $\text{log} \ (P_{\text{orb}} \ \text{(d)})$ & $T_{\mathrm{eff}} \ (\mathrm{K})$ & $e$ & $\omega$ (rad) & $p$ modes & References\\
 \hline
 \endfirsthead
 \caption{Continued.}\\
\hline
\hline
 KIC & $N_{\text{ind}}$ & $f_{\mathrm{max}}$ (d$^{-1}$) & $\text{log} \ (P_{\text{orb}} \ \text{(d)})$ & $T_{\mathrm{eff}} \ (\mathrm{K})$ & $e$ & $\omega$ (rad) & $p$ modes & References\\
\hline
\endhead
\hline
\insertTableNotes\\\\
\endfoot
\begin{filecontents*}{EB_sample.csv}
KIC,Nfreq,fmax,logP,Morph,Teff,e,omega,pmodes,refs
2720354,44,0.64896,0.450454,0.46,6513,0.0614,1.640,,\citetalias{Gaulme2019}
3127817,26,0.23569,0.636201,0.48,6504,0.0502,4.777,,
3248332,22,0.46456,0.867091,0.20,6578,0.1183,5.063,,
3327980,15,0.06892,0.626445,0.44,7321,0.0205,1.635,Present,\citetalias{Gaulme2019}
3352751,19,0.28099,0.543504,0.50,7909,0.0949,1.507,,
3544694,28,0.22904,0.584975,0.29,6172,0.0126,1.600,,
3867593$^{*,**}$,30,1.72913,1.865294,0.02,7037,,,,\citetalias{Debosscher2011,Gaulme2019,Li2020a}
4055765,26,0.88393,1.299551,0.10,6440,,,,\citetalias{Gaulme2019}
4067110,21,2.72055,0.584399,0.16,6499,,,,
4076952,13,0.08277,0.989502,0.38,6228,0.0448,1.159,,
4544587,37,2.01316,0.340265,0.49,8255,0.2765,5.738,,\citetalias{Hambleton2013,Gaulme2019}
4862625,21,0.37870,1.301035,0.06,6149,0.1693,3.106,,
4931073,6,0.86196,1.430579,0.08,6453,0.4128,3.295,,
4932691$^{*,**}$,42,0.98395,1.257968,0.10,7109,0.3797,3.215,,\citetalias{Kjurkchieva2016,Gaulme2019,Li2020a}
5024292,23,0.32312,0.633573,0.37,6147,0.0424,1.293,,
5217733,27,2.42189,2.207506,0.03,9116,0.6162,2.564,,\citetalias{Gaulme2019}
5384802,32,0.54996,0.784124,0.17,6203,0.0276,4.699,,
5565486$^{*,**}$,32,1.51263,0.451026,0.50,6471,0.0100,4.781,,\citetalias{Lurie2017,Gaulme2019,Li2020a}
5738698,16,0.15341,0.682034,0.41,6210,0.0546,4.723,,\citetalias{Matson2016}
5817566,12,0.70518,0.924909,0.47,7994,0.0448,4.724,,\citetalias{Gaulme2019}
5961350,36,0.34766,0.721206,0.14,6869,0.1433,4.706,,
6063448,21,0.01944,1.880915,0.03,6416,0.4612,6.055,Present,\citetalias{Lurie2017,Gaulme2019}
6109688,30,1.11643,1.148812,0.18,6845,0.3255,4.923,,\citetalias{Lurie2017,Gaulme2019}
6145939,36,0.58596,1.249086,0.05,6090,0.5566,1.560,,\citetalias{Lurie2017,Gaulme2019}
6147122,45,0.92076,1.188971,0.06,7625,0.2085,1.722,,\citetalias{Lurie2017}
6279974,17,0.62126,-0.094064,0.43,6022,0.5023,1.700,,
6290382$^{*,**}$,40,1.84985,0.787330,0.13,7016,0.6439,1.588,Present,\citetalias{Li2020a}
6362386,22,0.16419,0.662040,0.32,6983,0.0851,4.677,,
6449358,39,0.14915,0.761687,0.31,7449,0.1078,1.594,,
6523216,16,0.17204,1.155735,0.19,6200,0.1473,1.259,,
6631721,22,0.35052,1.137932,0.07,6153,,,,
6766748,28,0.12518,0.845276,0.25,6601,0.0199,4.537,,
6805146,19,0.52379,1.139240,0.16,6214,0.1492,5.623,,\citetalias{Gaulme2019}
6889235,23,0.16027,0.715056,0.43,9288,0.0079,4.733,,\citetalias{Rowe2010,Gaulme2019}
7025851,11,0.43586,0.670339,0.31,6054,0.0232,1.470,,
7107567,5,1.23518,-0.090124,0.50,6897,0.0047,1.860,,\citetalias{Bradley2015}
7422883,22,0.44354,1.057458,0.16,6639,0.0673,2.938,,\citetalias{Debosscher2011,Gaulme2019}
7449844,36,1.36898,0.061535,0.46,6834,0.0954,4.625,,
7599004,4,0.33479,0.683308,0.18,6118,0.0942,4.738,,\citetalias{Bradley2015}
7831363,10,0.72129,0.449827,0.25,6072,0.0614,1.525,,
7970760,19,0.11330,0.883628,0.25,6172,0.0228,1.498,,
8019043,26,0.49410,0.297889,0.34,6396,0.0922,4.765,,
8098728,10,0.44224,1.388900,0.35,6404,0.1501,2.484,,
8112013,11,0.65894,0.252979,0.47,6350,0.0083,4.506,,
8193315,42,1.46676,0.418943,0.43,6457,0.0052,1.188,Present,\citetalias{Gaulme2019}
8196180,32,0.28453,0.564863,0.31,7114,0.1511,3.140,,\citetalias{Kjurkchieva2018}
8197761$^{*,**}$,35,1.02452,1.295313,0.07,7068,,,,\citetalias{Sowicka2017,Li2019a,Gaulme2019,Li2020a}
8316503,22,0.19483,0.704608,0.25,6103,0.1097,6.220,,\citetalias{Kjurkchieva2016}
8488876,7,0.37705,0.763569,0.18,6957,0.1438,5.988,,
8504570,34,0.24792,0.602896,0.35,6874,0.0187,4.715,Present,\citetalias{Gaulme2019}
8560861,18,0.53515,1.504787,0.12,7647,0.0384,0.299,,\citetalias{Borkovits2014,Gaulme2019}
8569819$^{*,**}$,37,2.16310,1.319104,0.17,7137,0.4046,4.723,Present,\citetalias{Kurtz2015a,Gaulme2019,Li2020a}
8700506,15,0.45330,1.641452,0.04,6608,0.4835,0.968,,
8719419,27,0.74121,1.101243,0.07,6642,,,,\citetalias{Gaulme2019}
8823868,22,0.72297,1.377964,0.16,9751,0.0068,4.651,,\citetalias{Rowe2010,Gaulme2019}
8878681,13,0.23271,0.397303,0.33,6504,0.0657,4.690,,
9048145,33,0.19838,0.937910,0.15,6484,0.0056,1.573,,
9236858$^{**}$,32,1.78603,0.404334,0.45,6510,0.0545,1.569,Present,\citetalias{Gaulme2019,Li2020a}
9278021,14,0.62750,0.526271,0.19,6331,0.0270,0.900,,
9392702,28,0.22440,0.592101,0.37,6170,0.0247,4.687,,
9552608,34,1.32136,0.966171,0.11,7906,0.3194,5.544,,\citetalias{Lurie2017}
9637299,67,0.53663,0.274721,0.44,6061,0.0316,1.536,,
9711751,33,0.59909,0.233377,0.49,6429,0.0000,3.142,,
9850387$^{*,**}$,34,0.84868,0.439096,0.47,6808,0.0046,1.604,Present,\citetalias{Gaulme2019,Zhang2020,Li2020a}
9898364,55,1.47549,0.853235,0.11,7300,,,,
9911112,6,0.05405,0.368036,0.37,8750,0.0420,4.745,,
9936698,10,0.15046,0.756800,0.27,6393,0.0496,1.523,,\citetalias{Bradley2015}
10028352,3,1.71287,0.142014,0.36,6191,0.6149,1.603,,
10156064,12,0.19744,0.686273,0.32,7424,0.0310,1.532,,
10453521,8,0.34836,-0.636805,0.36,6541,,,,
10486425$^{**}$,45,1.31896,0.722207,0.25,7018,0.0109,1.494,,\citetalias{Alicavus2014,Zhang2018,Gaulme2019,Li2020a}
10489521,22,0.27982,0.508463,0.38,6147,0.0396,1.555,,
10549576,35,0.81913,0.958538,0.20,7492,0.0669,4.655,Present,\citetalias{Lurie2017}
10659313,24,0.13946,1.183251,0.06,6167,,,,
10686876,27,0.17696,0.418041,0.45,7944,0.0044,4.522,Present,
10920086,18,0.90127,0.506937,0.37,6478,0.8804,1.614,,
10937609,20,0.39576,0.410006,0.50,6016,0.0235,4.756,,
10987439,12,0.61580,1.028352,0.10,6182,0.0449,5.482,,\citetalias{Helminiak2019}
11021252,24,0.31271,0.514440,0.22,6165,0.0036,4.538,,
11099351,24,0.42547,0.376204,0.29,9353,0.0407,4.710,,
11231334,30,0.68783,0.618830,0.44,6278,0.0266,4.775,,\citetalias{Lurie2017}
11252617,31,0.44401,0.651096,0.19,6089,0.0349,1.508,,
11285625,7,0.56747,1.033039,0.23,6882,0.0293,4.685,,\citetalias{Debosscher2013,Gaulme2019}
11358392,20,1.01685,-0.006574,0.44,6195,0.0407,4.693,,
11359305,12,0.33875,1.454648,0.02,6101,0.4907,2.741,,
11671429,21,3.50450,2.051012,0.07,7363,0.4377,5.705,Present,\citetalias{Uytterhoeven2011,Gaulme2019}
11817750,25,0.52841,0.989563,0.15,6930,0.0391,1.392,,\citetalias{Lurie2017,Gaulme2019}
11820830$^{**}$,11,0.90456,1.104892,0.08,7007,,,,\citetalias{Gaulme2019,Li2020a}
11913071,31,0.22199,0.573780,0.37,8329,0.0035,2.120,,\citetalias{Kjurkchieva2017a}
11923819,20,1.84782,1.520608,0.07,7724,0.2912,2.670,,\citetalias{Lurie2017,Gaulme2019}
12167361,27,1.20897,1.680616,0.00,8017,,,,\citetalias{Lurie2017,Gaulme2019}
12405950,10,1.38396,0.548958,0.28,6808,0.1526,4.147,,
12470041$^{**}$,25,1.47773,1.166363,0.04,7290,,,,\citetalias{Gaulme2019,Li2020a}
\end{filecontents*}
\csvreader[head to column names]{EB_sample.csv}{}{\KIC & \Nfreq & \fmax & \logP & \Teff & \e & \omega & \pmodes & \refs\\}
\end{longtable}
}
\end{ThreePartTable}

\newpage
\begin{figure*}[!t]
\begin{center}
\includegraphics{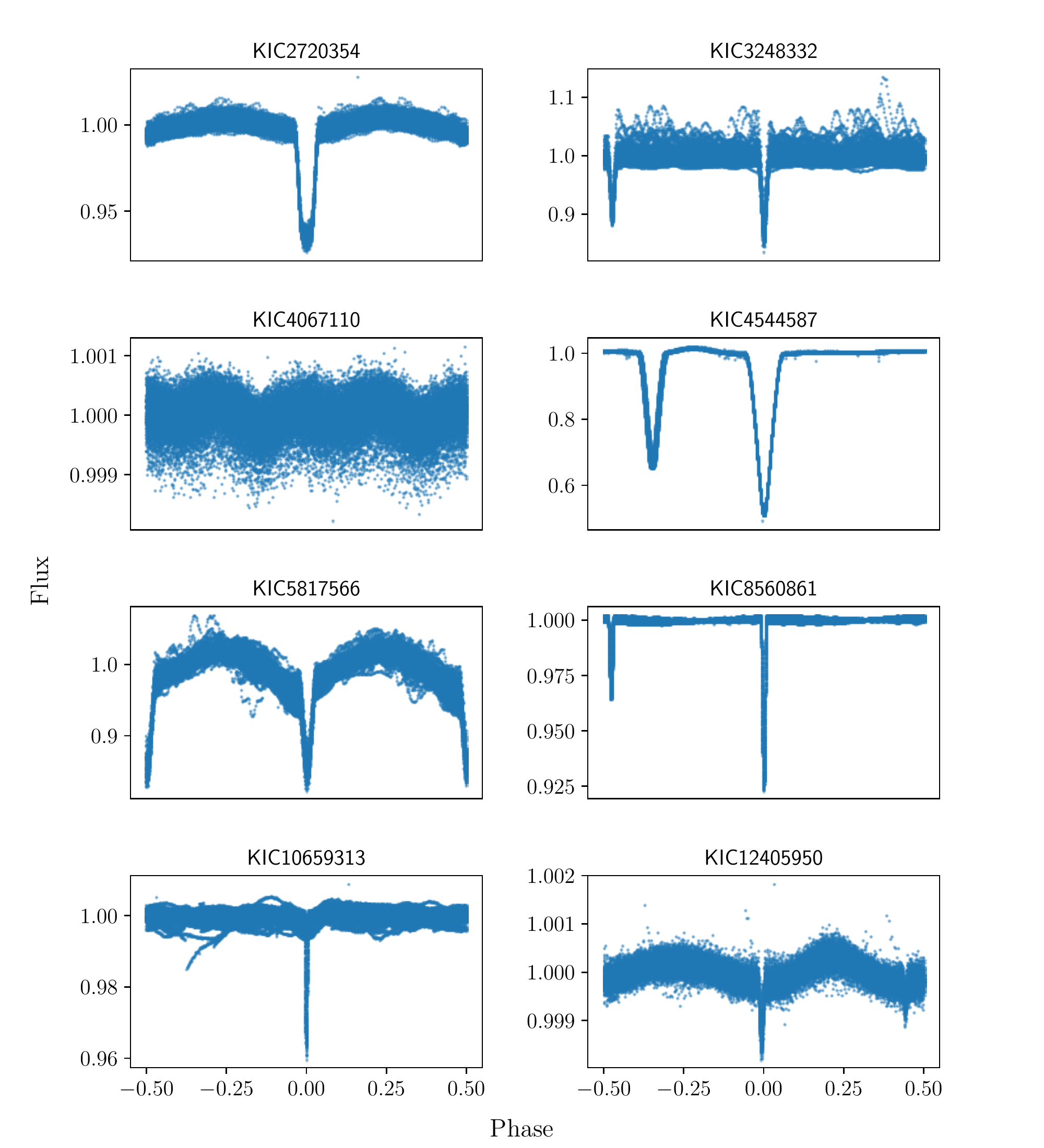}
\caption[Phase-folded light curves of eclipsing binary systems with $g$-mode pulsating components.]{Selection of phase-folded light curves of eclipsing binary systems with $g$-mode pulsating components, showing the plethora of variation in the visual morphologies of these systems.}
\label{fig: LC_selection}
\end{center}
\end{figure*}

\section{List of frequencies of KIC9850387}
\label{sec: freqlist}

\setlength{\tabcolsep}{12pt}
\renewcommand{\arraystretch}{1.3}
\LTcapwidth=\linewidth
\begin{longtable}{ccccccccc}
\caption{Full list of frequencies extracted from the light curve of KIC9850387 after the eclipses were removed, with their corresponding amplitudes, phases, and respective errors. The S/N values are based on the mean amplitude of the residual Lomb-Scargle periodogram in a 1 d$^{-1}$ window centred on each extracted frequency. Frequencies that form part of the $\ell=1$ period-spacing series are indicated with a * and frequencies that form part of the $\ell=2$ period-spacing series are indicated with a **.}\label{tab: freqlist}\\
\hline
\hline
 & $f$ (d$^{-1}$) & $\sigma_{\text{f}}$ (d$^{-1}$) & $A$ (ppm) & $\sigma_{\text{A}}$ (ppm) & $\phi$ ($2\pi$ rad) & $\sigma_{\phi}$ ($2\pi$ rad) & S/N & Combination\\
 \hline
 \endfirsthead
 \caption{Continued.}\\
\hline
\hline
 & $f$ (d$^{-1}$) & $\sigma_{\text{f}}$ (d$^{-1}$) & $A$ (ppm) & $\sigma_{\text{A}}$ (ppm) & $\phi$ ($2\pi$ rad) & $\sigma_{\phi}$ (rad) & S/N & Combination\\
\hline
\endhead
\hline
\endfoot
 \begin{filecontents*}{KIC9850387_frequencies.csv}
index,freq,freqerr,amp,amperr,phase,phaseerr,snr,swb,Combination
$f_{0}^{*}$,0.84868,0.00001,2087,74,$0.43$,$0.04$,500,153.0,
$f_{1}^{*}$,1.04635,0.00001,1848,68,$0.24$,$0.04$,425,142.1,
$f_{2}^{*}$,0.89709,0.00001,1835,61,$0.39$,$0.03$,405,135.5,
$f_{3}^{*}$,0.87225,0.00001,1715,55,$-0.47$,$0.03$,399,125.6,
$f_{4}^{*}$,1.01258,0.00001,1532,50,$-0.34$,$0.03$,328,116.7,
$f_{5}^{*}$,0.98020,0.00001,1298,47,$-0.14$,$0.04$,269,97.0,
$f_{6}^{*}$,0.92351,0.00001,1239,42,$0.32$,$0.03$,296,91.6,
$f_{7}^{*}$,1.08294,0.00002,770,40,$-0.14$,$0.05$,160,59.8,
$f_{8}^{*}$,0.95064,0.00002,758,39,$0.07$,$0.05$,144,56.3,
$f_{9}^{*}$,1.25773,0.00002,749,38,$-0.49$,$0.05$,160,63.6,
$f_{10}^{*,**}$,1.42837,0.00002,628,33,$-0.42$,$0.05$,137,55.8,
$f_{11}^{*}$,1.16400,0.00002,594,31,$0.27$,$0.05$,129,47.8,
$f_{12}^{**}$,1.15797,0.00002,503,30,$0.07$,$0.06$,110,40.3,
$f_{13}^{**}$,1.32428,0.00002,466,29,$0.24$,$0.06$,102,40.1,
$f_{14}$,0.21637,0.00002,464,29,$0.28$,$0.06$,105,33.5,2$f_{5}$-2$f_{3}$
$f_{15}$,13.21492,0.00003,451,31,$-0.22$,$0.07$,101,127.4,
$f_{16}^{*}$,0.80496,0.00002,448,29,$0.07$,$0.07$,101,32.3,
$f_{17}^{**}$,1.35729,0.00002,421,28,$0.48$,$0.07$,91,36.3,
$f_{18}^{*}$,0.82645,0.00002,401,27,$0.12$,$0.07$,82,29.1,
$f_{19}$,13.24803,0.00003,311,24,$0.18$,$0.08$,69,88.5,
$f_{20}$,0.46937,0.00003,305,24,$-0.03$,$0.08$,57,22.0,2$f_{7}$-2$f_{0}$
$f_{21}$,0.61696,0.00003,295,24,$-0.27$,$0.08$,63,21.5,2$f_{17}$-2$f_{1}$
$f_{22}$,5.33563,0.00003,278,24,$-0.17$,$0.08$,60,19.2,
$f_{23}$,5.67656,0.00004,236,22,$-0.34$,$0.09$,52,16.9,
$f_{24}$,0.38714,0.00004,227,22,$-0.47$,$0.10$,53,16.4,
$f_{25}$,0.35262,0.00004,223,22,$-0.10$,$0.10$,40,16.1,2$f_{1}$-2$f_{3}$
$f_{26}$,0.44874,0.00004,207,21,$-0.01$,$0.10$,47,14.9,
$f_{27}$,0.75417,0.00004,204,21,$-0.18$,$0.10$,46,14.7,2$f_{17}$-2$f_{5}$
$f_{28}^{**}$,1.39314,0.00004,203,21,$0.31$,$0.10$,41,17.8,
$f_{29}^{**}$,1.23595,0.00004,202,20,$0.34$,$0.10$,39,17.0,
$f_{30}$,0.44194,0.00004,200,20,$0.22$,$0.10$,44,14.4,2$f_{1}$-2$f_{18}$
$f_{31}^{*}$,1.36618,0.00004,197,20,$-0.37$,$0.10$,43,17.1,
$f_{32}^{**}$,1.26400,0.00004,193,20,$-0.19$,$0.10$,58,16.4,
$f_{33}$,0.28326,0.00004,179,19,$0.37$,$0.11$,41,12.9,
$f_{34}$,0.40108,0.00004,175,19,$0.38$,$0.11$,35,12.6,$f_{0}$-$f_{26}$
$f_{35}^{**}$,1.29420,0.00004,172,19,$-0.33$,$0.11$,22,14.8,
$f_{36}^{*,**}$,1.20883,0.00004,163,19,$-0.28$,$0.11$,35,13.5,
$f_{37}^{*}$,1.49669,0.00005,141,17,$-0.33$,$0.12$,31,12.7,
$f_{38}$,4.35484,0.00005,133,17,$-0.05$,$0.13$,33,12.4,
$f_{39}$,0.34365,0.00005,132,17,$0.20$,$0.13$,22,9.5,2$f_{1}$-2$f_{3}$
$f_{40}$,0.64895,0.00005,131,16,$0.08$,$0.12$,30,9.6,$f_{10}$-2$f_{24}$
$f_{41}$,1.55092,0.00005,131,16,$-0.36$,$0.13$,24,12.0,$f_{18}$+2$f_{\text{orb}}$
$f_{42}$,0.01943,0.00005,130,16,$-0.50$,$0.12$,18,9.4,$f_{0}$-$f_{18}$
$f_{43}$,4.60806,0.00005,126,16,$0.03$,$0.13$,34,10.4,2$f_{5}$+2$f_{13}$
$f_{44}$,0.69038,0.00005,126,16,$-0.42$,$0.13$,26,9.2,2$f_{10}$-2$f_{7}$
$f_{45}$,5.95992,0.00005,124,16,$0.21$,$0.13$,22,9.5,$f_{23}$+$f_{33}$
$f_{46}$,1.75005,0.00005,117,15,$0.35$,$0.13$,24,11.5,$f_{0}$+$f_{2}$
$f_{47}$,0.40704,0.00005,114,15,$-0.30$,$0.13$,25,8.2,
$f_{48}$,0.73581,0.00005,112,15,$0.43$,$0.13$,23,8.0,$f_{26}$+$f_{33}$
$f_{49}$,0.55089,0.00005,110,15,$0.13$,$0.13$,27,8.0,2$f_{13}$-2$f_{1}$
$f_{50}^{*}$,0.78498,0.00005,102,14,$-0.26$,$0.14$,22,7.4,
$f_{51}$,0.57052,0.00005,98,14,$0.27$,$0.14$,24,7.2,$f_{0}$-$f_{33}$
$f_{52}$,0.71932,0.00005,98,14,$-0.34$,$0.14$,18,7.1,
$f_{53}$,0.70087,0.00005,97,14,$-0.16$,$0.15$,18,7.0,$f_{10}$-2$f_{\text{orb}}$
$f_{54}$,0.02543,0.00006,96,14,$-0.11$,$0.15$,21,6.9,$f_{2}$-$f_{3}$
$f_{55}$,12.91505,0.00006,96,14,$-0.23$,$0.15$,19,27.5,
$f_{56}$,0.47562,0.00005,96,14,$-0.46$,$0.15$,21,6.9,$f_{6}$-$f_{26}$
$f_{57}$,4.69590,0.00006,94,14,$-0.49$,$0.15$,20,7.4,
$f_{58}$,13.79063,0.00006,93,14,$-0.16$,$0.15$,19,25.6,$f_{3}$+$f_{55}$
$f_{59}$,0.37398,0.00006,91,14,$0.09$,$0.15$,18,6.6,2$f_{7}$-2$f_{2}$
$f_{60}$,0.38166,0.00006,88,14,$-0.02$,$0.15$,20,6.4,$f_{18}$-$f_{26}$
$f_{61}$,4.92865,0.00006,88,14,$0.37$,$0.16$,20,6.7,$f_{22}$-$f_{47}$
$f_{62}$,5.20383,0.00006,88,14,$-0.47$,$0.16$,21,6.2,$f_{0}$+$f_{38}$
$f_{63}$,0.88059,0.00006,85,14,$-0.28$,$0.16$,15,6.2,2$f_{18}$-2$f_{24}$
$f_{64}$,0.46014,0.00006,83,14,$0.08$,$0.16$,19,6.0,$f_{0}$-$f_{24}$
$f_{65}^{**}$,1.18253,0.00006,82,13,$0.49$,$0.16$,19,6.7,
$f_{66}$,5.55356,0.00006,81,13,$0.03$,$0.17$,17,5.6,
$f_{67}$,0.54203,0.00006,80,13,$0.42$,$0.17$,17,5.8,$f_{18}$-$f_{33}$
$f_{68}$,5.23340,0.00007,75,13,$0.20$,$0.17$,17,5.3,
$f_{69}$,5.61907,0.00007,73,13,$0.11$,$0.18$,15,5.1,$f_{22}$+$f_{33}$
$f_{70}$,4.32303,0.00007,72,13,$0.19$,$0.18$,14,6.8,
$f_{71}$,1.59586,0.00007,70,13,$0.02$,$0.18$,18,6.6,$f_{3}$+2$f_{\text{orb}}$
$f_{72}$,0.93166,0.00007,70,13,$-0.21$,$0.18$,16,5.2,$f_{\text{orb}}$+2$f_{33}$
$f_{73}$,0.07876,0.00007,70,13,$0.09$,$0.18$,19,5.0,$f_{0}$-2$f_{24}$
$f_{74}$,6.43470,0.00007,70,13,$-0.24$,$0.18$,13,6.8,
$f_{75}$,0.74797,0.00007,69,13,$0.44$,$0.18$,14,4.9,2$f_{13}$-2$f_{8}$
$f_{76}$,5.11149,0.00007,69,13,$0.47$,$0.18$,14,4.9,$f_{23}$-2$f_{33}$
$f_{77}$,2.08505,0.00007,68,13,$-0.20$,$0.18$,10,7.3,$f_{17}$+2$f_{\text{orb}}$
$f_{78}$,0.97089,0.00007,67,13,$0.37$,$0.19$,15,5.0,2$f_{17}$-2$f_{3}$
$f_{79}$,1.70800,0.00007,67,12,$0.30$,$0.19$,12,6.4,$f_{5}$+2$f_{\text{orb}}$
$f_{80}^{*}$,1.12186,0.00007,66,12,$-0.40$,$0.19$,12,5.2,
$f_{81}$,0.05248,0.00007,66,12,$0.36$,$0.19$,13,4.7,2$f_{3}$-2$f_{0}$
$f_{82}$,5.30299,0.00007,66,12,$-0.01$,$0.19$,12,4.6,$f_{8}$+$f_{38}$
$f_{83}$,5.31218,0.00007,66,12,$0.48$,$0.19$,15,4.6,
$f_{84}$,6.06405,0.00007,65,12,$0.04$,$0.19$,15,5.3,$f_{22}$+2$f_{\text{orb}}$
$f_{85}$,0.71197,0.00007,65,12,$-0.23$,$0.19$,12,4.7,2$f_{52}$-2$f_{\text{orb}}$
$f_{86}$,6.25269,0.00007,65,12,$0.35$,$0.19$,15,5.7,$f_{38}$+2$f_{8}$
$f_{87}$,0.31524,0.00007,65,12,$-0.03$,$0.19$,14,4.7,$f_{1}$-2$f_{\text{orb}}$
$f_{88}$,0.04131,0.00007,65,12,$-0.27$,$0.19$,12,4.7,$f_{3}$-$f_{18}$
$f_{89}$,5.22041,0.00007,63,12,$0.34$,$0.19$,14,4.4,$f_{70}$+2$f_{26}$
$f_{90}$,0.94220,0.00007,62,12,$0.33$,$0.19$,16,4.6,$f_{13}$-$f_{24}$
$f_{91}$,1.99314,0.00007,62,12,$-0.33$,$0.19$,14,6.6,$f_{1}$+$f_{8}$
$f_{92}$,0.68472,0.00007,62,12,$-0.15$,$0.19$,15,4.5,2$f_{13}$-2$f_{5}$
$f_{93}$,0.06316,0.00007,62,12,$0.06$,$0.19$,14,4.4,$f_{1}$-$f_{5}$
$f_{94}$,6.12118,0.00007,61,12,$-0.25$,$0.19$,15,5.1,$f_{23}$+$f_{26}$
$f_{95}$,4.94744,0.00007,61,12,$0.39$,$0.19$,11,4.6,2$f_{1}$+2$f_{10}$
$f_{96}$,6.33135,0.00007,61,12,$-0.38$,$0.19$,11,5.6,$f_{66}$+2$f_{24}$
$f_{97}^{*}$,1.30961,0.00007,60,12,$-0.31$,$0.19$,14,5.2,
$f_{98}$,0.83844,0.00007,60,12,$-0.27$,$0.19$,13,4.4,$f_{24}$+$f_{26}$
$f_{99}$,5.17190,0.00007,60,12,$-0.31$,$0.20$,13,4.2,$f_{38}$+2$f_{47}$
$f_{100}$,0.90788,0.00007,60,12,$0.04$,$0.19$,13,4.4,2$f_{13}$-2$f_{3}$
$f_{101}$,5.53950,0.00007,59,12,$-0.33$,$0.20$,11,4.1,$f_{0}$+$f_{57}$
$f_{102}$,5.56443,0.00007,59,12,$0.44$,$0.20$,13,4.1,$f_{3}$+$f_{57}$
$f_{103}$,1.74000,0.00007,58,12,$-0.07$,$0.20$,13,5.8,2$f_{3}$
$f_{104}$,0.03488,0.00008,57,12,$0.34$,$0.20$,12,4.1,$f_{5}$-$f_{8}$
$f_{105}$,1.62482,0.00008,57,12,$0.13$,$0.20$,11,5.4,$f_{2}$+2$f_{\text{orb}}$
$f_{106}$,6.15122,0.00008,57,12,$-0.30$,$0.20$,13,4.8,$f_{38}$+2$f_{2}$
$f_{107}$,1.46368,0.00008,56,11,$0.03$,$0.20$,13,5.0,2$f_{26}$+2$f_{33}$
$f_{108}$,5.68552,0.00008,56,11,$0.29$,$0.20$,13,4.0,$f_{26}$+$f_{68}$
$f_{109}$,6.34719,0.00008,56,11,$-0.15$,$0.20$,13,5.2,$f_{57}$+2$f_{18}$
$f_{110}$,1.77363,0.00008,56,11,$-0.37$,$0.20$,10,5.6,$f_{1}$+2$f_{\text{orb}}$
$f_{111}$,0.06921,0.00008,56,11,$-0.05$,$0.20$,13,4.0,$f_{1}$-$f_{5}$
$f_{112}$,5.77751,0.00008,56,11,$0.24$,$0.20$,15,4.0,$f_{7}$+$f_{57}$
$f_{113}$,2.19134,0.00008,55,11,$-0.11$,$0.20$,12,6.0,$f_{3}$+$f_{13}$
$f_{114}$,1.44703,0.00008,55,11,$-0.06$,$0.20$,11,4.9,$f_{\text{orb}}$+$f_{7}$
$f_{115}$,1.93149,0.00008,55,11,$0.46$,$0.21$,13,5.7,$f_{0}$+$f_{7}$
$f_{116}$,2.90235,0.00008,55,11,$0.05$,$0.21$,11,7.9,2$f_{1}$+2$f_{47}$
$f_{117}$,3.63001,0.00008,55,11,$-0.07$,$0.21$,13,7.6,$f_{5}$+2$f_{13}$
$f_{118}$,0.22379,0.00008,54,11,$-0.07$,$0.21$,11,3.9,$f_{8}$-2$f_{\text{orb}}$
$f_{119}$,5.43336,0.00008,54,11,$-0.32$,$0.21$,11,3.7,$f_{7}$+$f_{38}$
$f_{120}$,1.86088,0.00008,54,11,$-0.40$,$0.21$,10,5.4,$f_{15}$-2$f_{23}$
$f_{121}$,1.89496,0.00008,53,11,$-0.19$,$0.21$,12,5.4,$f_{0}$+$f_{1}$
$f_{122}$,2.05191,0.00008,53,11,$0.12$,$0.21$,12,5.6,$f_{13}$+2$f_{\text{orb}}$
$f_{123}$,5.46795,0.00008,53,11,$-0.35$,$0.21$,11,3.7,$f_{57}$+2$f_{24}$
$f_{124}$,5.72186,0.00008,53,11,$0.14$,$0.21$,9,3.8,$f_{22}$+$f_{24}$
$f_{125}$,6.45090,0.00008,52,11,$0.05$,$0.21$,10,5.1,$f_{23}$+2$f_{24}$
$f_{126}$,0.33632,0.00008,52,11,$0.47$,$0.21$,10,3.8,2$f_{26}$-2$f_{33}$
$f_{127}$,13.12638,0.00008,52,11,$0.35$,$0.21$,12,14.7,
$f_{128}$,3.02853,0.00008,51,11,$0.06$,$0.22$,9,7.5,$f_{23}$-2$f_{13}$
$f_{129}$,5.19785,0.00008,51,11,$-0.01$,$0.22$,11,3.6,$f_{3}$+$f_{70}$
$f_{130}$,11.15525,0.00008,50,11,$0.47$,$0.22$,11,13.3,$f_{19}$-2$f_{1}$
$f_{131}$,5.64499,0.00008,50,11,$-0.33$,$0.22$,11,3.6,$f_{8}$+$f_{57}$
$f_{132}$,5.66209,0.00008,50,11,$0.34$,$0.22$,11,3.6,$f_{74}$-2$f_{24}$
$f_{133}$,2.17463,0.00008,50,11,$-0.19$,$0.22$,11,5.4,$f_{0}$+$f_{13}$
$f_{134}$,12.32594,0.00008,50,11,$0.01$,$0.22$,11,13.5,2$f_{18}$+2$f_{22}$
$f_{135}$,5.39831,0.00008,50,11,$-0.11$,$0.22$,12,3.5,$f_{23}$-$f_{33}$
$f_{136}$,5.84119,0.00008,50,11,$-0.16$,$0.22$,11,3.7,$f_{33}$+$f_{66}$
$f_{137}$,0.60009,0.00008,50,11,$-0.01$,$0.22$,10,3.6,$f_{13}$-2$f_{\text{orb}}$
$f_{138}$,5.70183,0.00008,50,11,$0.42$,$0.22$,11,3.6,$f_{\text{orb}}$+$f_{22}$
$f_{139}$,0.14024,0.00008,50,11,$0.46$,$0.22$,11,3.6,$f_{3}$-2$f_{\text{orb}}$
$f_{140}$,4.53621,0.00008,49,11,$0.17$,$0.22$,11,4.2,$f_{19}$-2$f_{38}$
$f_{141}$,5.32024,0.00008,49,11,$0.23$,$0.22$,11,3.4,
$f_{142}$,1.06716,0.00008,49,11,$0.46$,$0.22$,11,3.8,2$f_{2}$-2$f_{\text{orb}}$
$f_{143}$,6.26639,0.00008,49,11,$-0.32$,$0.22$,10,4.3,$f_{57}$+2$f_{50}$
$f_{144}$,0.10660,0.00008,48,11,$-0.15$,$0.22$,10,3.5,2$f_{6}$-2$f_{3}$
$f_{145}$,5.96895,0.00008,48,11,$0.15$,$0.22$,10,3.7,
$f_{146}$,5.63860,0.00008,48,11,$0.32$,$0.22$,10,3.4,$f_{47}$+$f_{68}$
$f_{147}$,6.40452,0.00008,47,11,$-0.15$,$0.23$,10,4.6,$f_{23}$+2$f_{\text{orb}}$
$f_{148}$,0.79503,0.00008,47,11,$-0.34$,$0.23$,10,3.4,2$f_{0}$-2$f_{26}$
$f_{149}$,1.69724,0.00009,46,11,$-0.46$,$0.23$,10,4.4,2$f_{0}$
$f_{150}$,0.66479,0.00009,46,11,$0.34$,$0.23$,10,3.4,$f_{8}$-$f_{33}$
$f_{151}$,1.19574,0.00009,46,11,$-0.21$,$0.23$,10,3.8,2$f_{1}$-2$f_{26}$
$f_{152}$,2.43255,0.00009,46,11,$-0.23$,$0.23$,9,5.4,$f_{50}$+2$f_{18}$
$f_{153}$,4.66412,0.00009,46,11,$0.34$,$0.23$,10,3.7,$f_{68}$-2$f_{33}$
$f_{154}$,5.92386,0.00009,46,10,$0.49$,$0.23$,9,3.4,$f_{38}$+2$f_{50}$
$f_{155}$,4.91186,0.00009,46,10,$-0.14$,$0.23$,10,3.5,
$f_{156}$,4.82793,0.00009,46,11,$0.47$,$0.23$,10,3.5,$f_{66}$-2$f_{\text{orb}}$
$f_{157}$,1.72093,0.00009,45,10,$0.42$,$0.23$,10,4.4,$f_{0}$+$f_{3}$
$f_{158}$,5.26077,0.00009,45,10,$0.10$,$0.23$,10,3.1,$f_{57}$+2$f_{33}$
$f_{159}$,6.46791,0.00009,45,10,$0.49$,$0.23$,10,4.4,
$f_{160}$,0.58241,0.00009,45,10,$0.15$,$0.23$,9,3.3,$f_{17}$-2$f_{24}$
$f_{161}$,5.08356,0.00009,45,10,$0.42$,$0.23$,10,3.2,$f_{38}$+2$f_{\text{orb}}$
$f_{162}$,0.74241,0.00009,44,10,$0.35$,$0.24$,10,3.2,2$f_{13}$-2$f_{8}$
$f_{163}$,4.58465,0.00009,44,10,$-0.40$,$0.24$,9,3.7,$f_{74}$-2$f_{6}$
$f_{164}$,5.58132,0.00009,44,10,$0.12$,$0.24$,10,3.0,
$f_{165}$,0.50359,0.00009,44,10,$-0.17$,$0.24$,10,3.2,$f_{8}$-$f_{26}$
$f_{166}$,1.90959,0.00009,44,10,$-0.40$,$0.24$,10,4.4,$f_{7}$+$f_{18}$
$f_{167}$,1.22543,0.00009,43,10,$-0.00$,$0.24$,9,3.6,$f_{26}$+2$f_{24}$
$f_{168}$,1.07624,0.00009,43,10,$0.25$,$0.24$,10,3.4,2$f_{6}$-2$f_{24}$
$f_{169}$,0.09365,0.00009,43,10,$-0.28$,$0.24$,9,3.1,2$f_{2}$-2$f_{0}$
$f_{170}$,3.31189,0.00009,43,10,$-0.04$,$0.24$,9,6.3,2$f_{3}$+2$f_{50}$
$f_{171}$,5.67001,0.00009,43,10,$0.10$,$0.24$,9,3.1,
$f_{172}$,0.53353,0.00009,43,10,$0.28$,$0.24$,9,3.1,$f_{17}$-$f_{18}$
$f_{173}$,0.41839,0.00009,43,10,$-0.05$,$0.24$,9,3.1,2$f_{7}$-2$f_{3}$
$f_{174}$,0.81838,0.00009,42,10,$-0.24$,$0.24$,9,3.1,2$f_{17}$-2$f_{8}$
$f_{175}$,5.27044,0.00009,42,10,$-0.05$,$0.24$,9,3.0,$f_{23}$-$f_{47}$
$f_{176}$,1.57607,0.00009,42,10,$0.11$,$0.24$,9,3.9,$f_{0}$+2$f_{\text{orb}}$
$f_{177}$,6.73437,0.00009,42,10,$-0.45$,$0.24$,9,5.0,
$f_{178}$,0.76895,0.00009,42,10,$-0.06$,$0.24$,9,3.0,2$f_{10}$-2$f_{1}$
$f_{179}$,2.55223,0.00009,42,10,$-0.19$,$0.24$,9,5.2,$f_{2}$+2$f_{18}$
$f_{180}$,4.48966,0.00009,42,10,$0.11$,$0.24$,9,3.6,
$f_{181}^{**}$,1.13428,0.00009,42,10,$-0.03$,$0.24$,9,3.3,
$f_{182}$,5.50452,0.00009,42,10,$0.11$,$0.24$,9,2.9,2$f_{10}$+2$f_{13}$
$f_{183}$,6.03286,0.00009,42,10,$-0.48$,$0.24$,9,3.3,$f_{52}$+$f_{83}$
$f_{184}$,0.32080,0.00009,42,10,$0.24$,$0.25$,9,3.0,$f_{1}$-2$f_{\text{orb}}$
$f_{185}$,6.37979,0.00009,42,10,$0.17$,$0.24$,10,3.9,$f_{1}$+$f_{22}$
$f_{186}$,5.69391,0.00009,42,10,$0.04$,$0.24$,9,3.0,$f_{50}$+$f_{155}$
$f_{187}$,4.80445,0.00009,42,10,$0.09$,$0.24$,9,3.2,2$f_{1}$+2$f_{17}$
$f_{188}$,4.37399,0.00009,41,10,$0.16$,$0.24$,9,3.8,$f_{159}$-2$f_{1}$
$f_{189}$,1.88481,0.00009,41,10,$0.04$,$0.25$,9,4.2,$f_{26}$+2$f_{52}$
$f_{190}$,1.10158,0.00009,41,10,$-0.46$,$0.25$,9,3.2,$f_{33}$+2$f_{47}$
$f_{191}$,1.14621,0.00009,41,10,$0.16$,$0.25$,9,3.3,$f_{10}$-$f_{33}$
$f_{192}$,6.49571,0.00009,41,10,$-0.32$,$0.25$,9,4.2,$f_{23}$+2$f_{47}$
\end{filecontents*}
\csvreader[head to column names]{KIC9850387_frequencies.csv}{}{\index & \freq & \freqerr & \amp & \amperr & \phase & \phaseerr & \swb & \Combination\\}
\end{longtable}
\twocolumn
\vspace{-10pt}

\end{appendix}

\end{document}